\documentclass[%
reprint,
superscriptaddress,
amsmath,amssymb,
pre,
]{revtex4-2}
\usepackage{subcaption}
\usepackage[labelfont=bf]{caption}
\usepackage[percent]{overpic}
\usepackage{xcolor}
\usepackage{graphicx}
\usepackage{dcolumn}
\usepackage{bm}
\usepackage[normalem]{ulem} 

\begin{document}
	
	
	\title{Channel Flow Around an Obstacle in a Two-Dimensional Soft Particles}
	
	\author{Bahaa Mazloum}
	\affiliation{Univ. Grenoble Alpes, CNRS, LIPhy, 38000 Grenoble, France}
	
	\author{Alexandre Stepanetz}
	\affiliation{Univ. Grenoble Alpes, CNRS, LIPhy, 38000 Grenoble, France}
	
	\author{Benjamin Dollet}
	\affiliation{Univ. Grenoble Alpes, CNRS, LIPhy, 38000 Grenoble, France}
	
	\author{Misaki Ozawa}
	\affiliation{Univ. Grenoble Alpes, CNRS, LIPhy, 38000 Grenoble, France}

	\date{\today}
	
	\begin{abstract}
		
		We numerically study confined channel flow around an obstacle using a two-dimensional soft particle model, inspired by experiments performed in the same geometry. We systematically vary the polydispersity, the external driving force, and the packing fraction of the system. Our simulations capture a broad range of plastic flow phenomenologies, from highly directional, sliding-like motion characteristic of crystalline materials to more isotropic and localized rearrangements typical of amorphous systems. We identify a threshold value of polydispersity that marks the crossover between crystalline-like and amorphous-like plasticity. In addition, we observe the existence of a critical external force, associated with the phenomenon of yield drag, above which the system reaches steady flow and below which it remains arrested. We determine a critical packing fraction above which such yield-drag behavior emerges. Our results provide a comprehensive framework for understanding the interplay between disorder, driving, and the presence of an obstacle in channel flows.
		
	\end{abstract}

	\maketitle
	
	
	\section{Introduction}

	The flow of particulate materials, including foams, emulsions, dense suspensions, soft colloidal systems, and granular materials, constitutes an important topic in soft matter physics~\cite{bonn2017yield}. These systems commonly exhibit complex and heterogeneous dynamics associated with intermittent plastic rearrangements~\cite{berthier2011dynamical}. The spatial organization of such rearrangements depends strongly on the degree of structural disorder. Weakly polydisperse, crystalline-like systems can display directional, sliding-like, and highly anisotropic motion~\cite{sethna2017deformation,Bragg1947,Oswald2014}, whereas strongly disordered amorphous systems generally exhibit more localized and isotropic rearrangements~\cite{Argon1979,Bulatov1994,nicolas2018deformation,richard2020predicting}. In dry liquid foams, for example, plastic deformation proceeds through elementary topological rearrangements known as T1 events~\cite{weaire1984soap}, in which neighboring bubbles exchange their relative positions.
	
	Much of our current understanding of the rheology and plasticity of particulate materials has been obtained using spatially homogeneous deformation protocols~\cite{bonn2017yield}. Typical examples include shear experiments performed using cone-and-plate or parallel-plate rheometers and numerical simulations employing periodic boundary conditions. Although these conventional protocols provide controlled conditions for characterizing bulk mechanical response, alternative loading geometries can reveal additional aspects of heterogeneous deformation and may be more directly relevant to practical transport processes.
	
	Channel flow provides one such complementary geometry. It is of considerable fundamental interest for investigating the rheology, plasticity, jamming, and flow heterogeneity of driven particulate materials~\cite{Katgert2008,campbell2010jamming,chen2012topological}. Channel flows are also relevant to numerous practical applications~\cite{isa2009velocity,Stevenson2012}, including the transport of particulate and soft materials through porous media, constrictions, and other confined geometries~\cite{Ma2012,Geraud2016}.
	
	The introduction of an obstacle into a channel provides an additional source of geometric heterogeneity. Flow around an obstacle is a classical configuration in fluid mechanics, exemplified by Stokes flow, and offers a useful geometry for probing the deformation and rheological response of complex materials. The obstacle produces a strongly heterogeneous stress and velocity field and can therefore induce plastic rearrangements. Such geometries have been investigated experimentally in particular, flowing around circular obstacles~\cite{chehata2003dense,dollet2007two,tokpavi2009experimental,Viitanen2019}.
	
	
	Despite its practical relevance and fundamental interest, channel flow around a circular obstacle has not been systematically investigated using particle-resolved models of dense soft materials. Most previous numerical studies of this geometry have instead relied on mesoscopic or continuum descriptions~\cite{zisis2002viscoplastic,mousavi2025elasto}.
	In this work, we develop an overdamped particle-based simulation model inspired by the experiments discussed above, particularly the foam-flow experiments performed by Dollet and Graner~\cite{dollet2007two}, and systematically explore how key system parameters control the flow behavior. In particular, we vary the degree of polydispersity $\delta$, the magnitude of the external driving force $f^{\rm ext}$, and the packing fraction $\phi$.

	In the first part of this study, we systematically investigate how $\delta$ influences heterogeneous plasticity under low driving $f^{\rm ext}$ and high packing fraction $\phi$. 
	We carefully analyze how the spatial pattern of plastic heterogeneity evolves, from directional, sliding-like motion in crystalline-like systems to more isotropic rearrangements in amorphous systems, and quantitatively evaluate the magnitude of anisotropy. 
	
	In the second part of this study, we focus on amorphous systems and examine how varying the magnitude of the external driving force modifies the flow behavior, ranging from heterogeneous dynamics at weak driving to more streamlined motion at stronger driving. Interestingly, at low driving, we observe the existence of a critical external force below which the system ceases to flow and remains in a jammed state, and above which it reaches a steady flowing state. This behavior is reminiscent of the yield-drag phenomenon reported in experiments and other types of simulations~\cite{Raufaste2007,Cantat2006}, as well as the yielding transition observed in amorphous materials under shear deformation protocols~\cite{nicolas2018deformation,bonn2017yield} and the physics of depinning transitions~\cite{fisher1998collective,reichhardt2016depinning}.
	Our simulation results also suggest the existence of a critical volume fraction above which such a threshold force appears, and below which the system always flows irrespective of the magnitude of the external driving.
	
	Although our simulation model is sufficiently generic to describe a broad class of soft-particle systems, we discuss in detail its relevance as a foam model, since the present study was originally motivated by foam experiments.
	
	The paper is organized as follows. 
	Section~\ref{sec:methods} describes the simulation methods. 
	In Section~\ref{sec:polydispersity}, we present results for varying polydispersity and examine how it influences the spatial organization of plastic dynamics. 
	Section~\ref{sec:external_driving} focuses on the effect of the magnitude of the external driving force on the flow behavior. 
	We then discuss the relevance of our results to foam systems in Sec.~\ref{sec:foam}.
	Finally, conclusions and discussions are provided in Section~\ref{sec:conclusion}.

	\section{Simulation methods}
	\label{sec:methods}

\begin{figure}
		\includegraphics[width=0.9\linewidth]{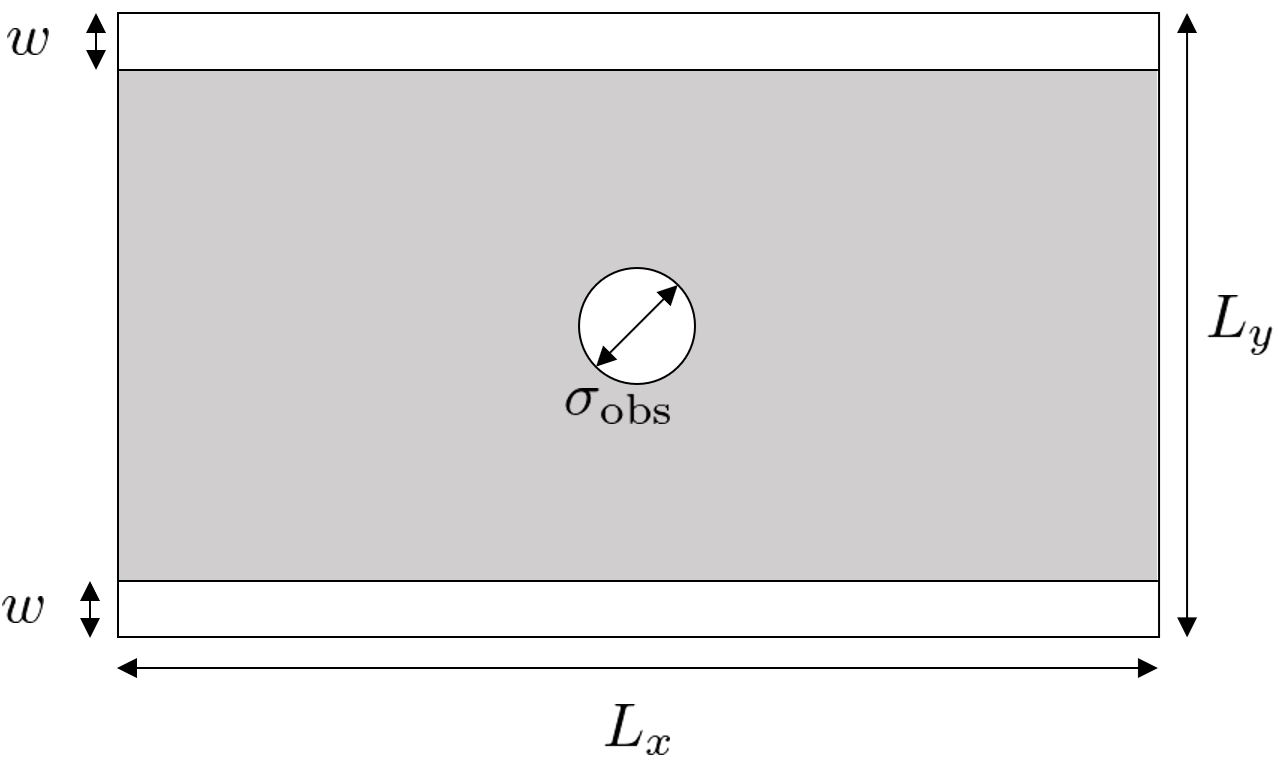}
		\captionsetup{justification=raggedright,singlelinecheck=false}
		\caption{Geometry of the two-dimensional rectangular simulation box. 
			Particles interact with the top and bottom walls, each of width $w$, 
			and with a central circular obstacle of diameter $\sigma_{\rm obs}$.
		}
		\label{fig:geometry}
	\end{figure}

	We construct a molecular dynamics simulation model of a two-dimensional soft particles flowing through a channel containing a circular obstacle at the center. The simulation geometry is illustrated in Fig.~\ref{fig:geometry}. The system consists of a rectangular simulation box with dimensions $L_x \times L_y$, confined by walls at the top and bottom. Each wall has a thickness $w=0.5$. A circular obstacle of diameter $\sigma_{\rm obs}=10$ is placed at the center of the channel. Periodic boundary conditions are applied along the $x$-direction. Throughout this study, we set $L_x = \gamma L_y$ with $\gamma = 3$. The value of $L_y$ depends on the packing fraction of the system (see below). In the present study, we use values in the range $L_y \approx 15.1 - 17.6$, depending on the packing fraction (see below).

	The system is represented by $N = 900$ soft, polydisperse disks. The degree of polydispersity is quantified by
	\begin{equation}
		\delta = \frac{\sqrt{\overline{\sigma^2} - \overline{\sigma}^2}}{\overline{\sigma}},
	\end{equation}
	where $\overline{\sigma}$ denotes the mean particle diameter, i.e., $\overline{\sigma}=\frac{1}{N}\sum_{i=1}^N \sigma_i$.
	
	We consider overdamped dynamics~\cite{durian1995foam} for the position of the $i$-th particle, ${\bf r}_i(t) = (x_i(t), y_i(t))$, which evolves according to
	\begin{equation}
		\zeta \frac{d {\bf r}_i(t)}{d t} =  {\bf f}_i^{\rm int} + {\bf f}_i^{\rm wall} + {\bf f}_i^{\rm obs}  + {\bf f}^{\rm ext} ,
		\label{eq:overdamped_dynamics}
	\end{equation}
	where ${\bf f}_i^{\rm int}$ is the interaction force from neighboring particles, ${\bf f}_i^{\rm wall}$ and ${\bf f}_i^{\rm obs}$ are the repulsive forces from the confining walls and the obstacle, respectively, and ${\bf f}^{\rm ext}$ denotes the external driving force.
	Here, $\zeta$ is the viscous damping coefficient, which we set to $\zeta = 1$ throughout this study.
	We integrate Eq.~(\ref{eq:overdamped_dynamics}) using the Euler method with a time step of $dt = 0.1$.

	The interparticle force ${\bf f}_i^{\rm int}$ derives from a pairwise purely repulsive potential acting only for overlapping particles ($r_{ij} < \sigma_{ij}$):
	\begin{equation} \label{Eq:potential}
		v_{ij}^{\rm int}(r_{ij}) = \frac{\epsilon}{\alpha}\left( 1 - \frac{r_{ij}}{\sigma_{ij}} \right)^{\alpha},
	\end{equation}
	where $r_{ij}=|{\bf r}_i-{\bf r}_j|$ and $\sigma_{ij} = (\sigma_i + \sigma_j)/2$. For $r_{ij} \geq \sigma_{ij}$, the interaction vanishes, i.e., $v_{ij}^{\rm int}(r_{ij}) = 0$. In this work, we report results obtained using a Hertzian contact interaction with exponent $\alpha = 5/2$. We have verified, however, that qualitatively the same behavior is observed for a harmonic potential with $\alpha = 2$.
	
	In the absence of thermal fluctuations, the natural microscopic timescale of the system is given by $t_0 =\zeta\, \overline{\sigma}^2/\epsilon$. 
	Throughout this work, we measure length, time, and energy in units of $\overline{\sigma}$, $t_0$, and $\epsilon$, respectively.
	
	For the wall confinement, we introduce a repulsive interaction between particles and the walls. 
	We choose the origin of coordinates at the bottom-left corner of the simulation box. 
	A particle with vertical position $y_i$ interacts with the bottom wall through a harmonic potential
	\begin{equation}
		V_{\rm bottom}^{\rm wall}(y_i) = 
		\begin{cases}
			\dfrac{K}{2} (y_i - w)^2, & 0 \le y_i \le w, \\[5mm]
			0, & \text{otherwise} ,
		\end{cases}
	\end{equation}
	which produces ${\bf f}_i^{\rm wall}=(0,K(w-y_i))$.
	The parameter $K$ controls the stiffness of the wall repulsion; in this study, we set a sufficiently large value, $K = 10$.
	The interaction with the top wall is set in the same way.
	
	The interaction with the central obstacle is likewise modeled by a harmonic repulsion. 
	Let $r_i$ be the distance between the $i$-th particle and the center of the obstacle, which is located at $(L_x/2,\, L_y/2)$:
	For $r_i < \sigma_{\rm obs}/2$, the particle experiences the repulsive potential
	\begin{equation}
		V^{\rm obs}(r_i) = \frac{K}{2}\left( r_i - \frac{\sigma_{\rm obs}}{2} \right)^2,
	\end{equation}
	while $V^{\rm obs}(r_i) = 0$ otherwise. 
	The resulting force is 
	\begin{equation}
		{\bf f}_i^{\rm obs} = -K\left(r_i - \frac{\sigma_{\rm obs}}{2}\right)
		\frac{(x_i - L_x/2,\, y_i - L_y/2)}{r_i}.
	\end{equation}
	
	For the external driving, we apply a constant force uniformly to all particles along the $x$-direction. 
	Namely, we set
	\begin{equation}
		{\bf f}^{\rm ext} = (f^{\rm ext},\, 0),
	\end{equation}
	and we systematically vary the magnitude $f^{\rm ext}$ to control the flow rate.

	We control the packing fraction of the system by considering only the area accessible to the particles, 
	i.e., the region outside the walls and the obstacle (gray shaded region in Fig.~\ref{fig:geometry}). 
	Accordingly, the packing fraction is defined as
	\begin{equation}
		\phi = 
		\frac{\displaystyle \sum_{i=1}^{N} \pi \left(\frac{\sigma_i}{2}\right)^2}
		{\,L_x (L_y - 2w) - \pi \left(\frac{\sigma_{\rm obs}}{2}\right)^2\, } .
	\end{equation}

	We start from a random particle configuration generated according to a Poisson process as the initial condition. The system then evolves according to the overdamped dynamics given by Eq.~(\ref{eq:overdamped_dynamics}). During the initial stage, the system exhibits a transient flow regime characterized by a decrease in potential energy. After a sufficiently long time, the system reaches a steady state in which the potential energy fluctuates around a constant value.
	In this paper, all physical observables and data are reported in the steady state unless otherwise stated.

	\section{Results: Effect of Polydispersity}
	\label{sec:polydispersity}

	We first investigate the effect of polydispersity under weak external driving,
	$f^{\rm ext} = 10^{-3}$, where particle rearrangements exhibit intermittent plastic
	behavior in a highly jammed system with packing fraction $\phi = 1.2$.
	
	\subsection{Low polydispersity: Crystalline system}

	\begin{figure*}[htbp]
		\centering
		\begin{subfigure}[t]{0.32\linewidth}
			\captionsetup{justification=raggedright, singlelinecheck=false, position=above}  
			\caption{$\Delta {\bf r}$ for $\Delta t=100$}
			\includegraphics[width=\linewidth]{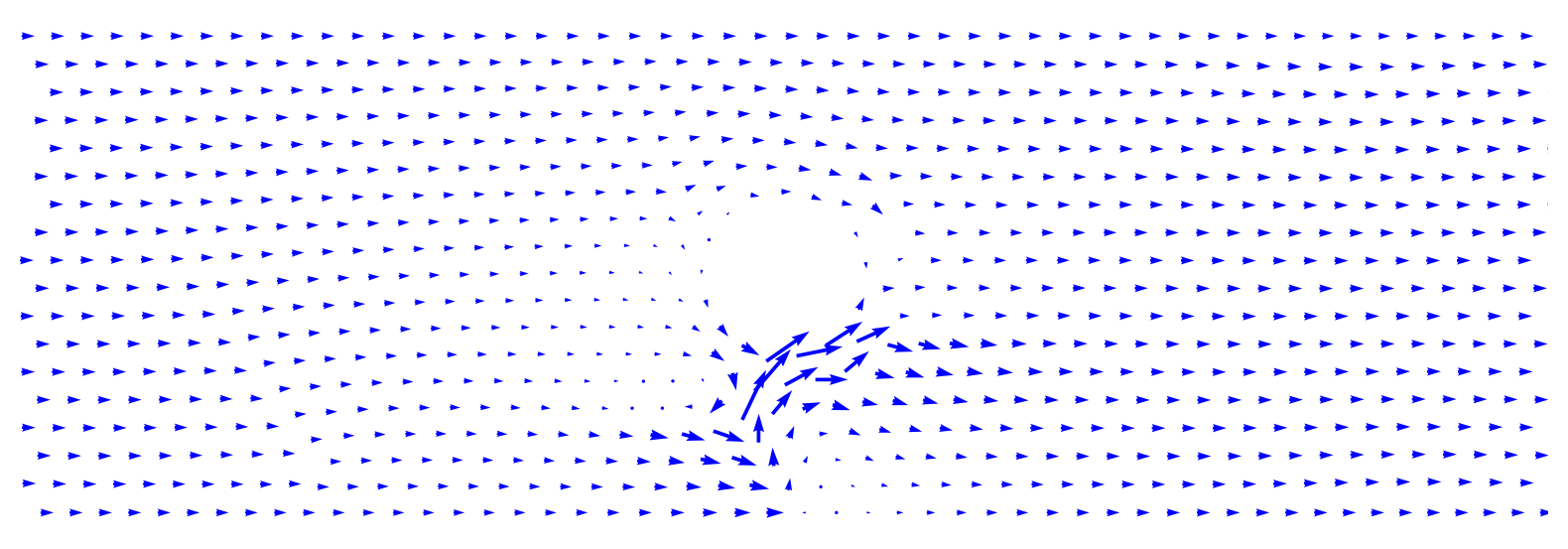}
			\label{fig:delta10a}
		\end{subfigure}
		\hfill
		\begin{subfigure}[t]{0.32\linewidth}
			\captionsetup{justification=raggedright, singlelinecheck=false, position=above}
			\caption{$\Delta {\bf r}$ for $\Delta t=300$}
			\includegraphics[width=\linewidth]{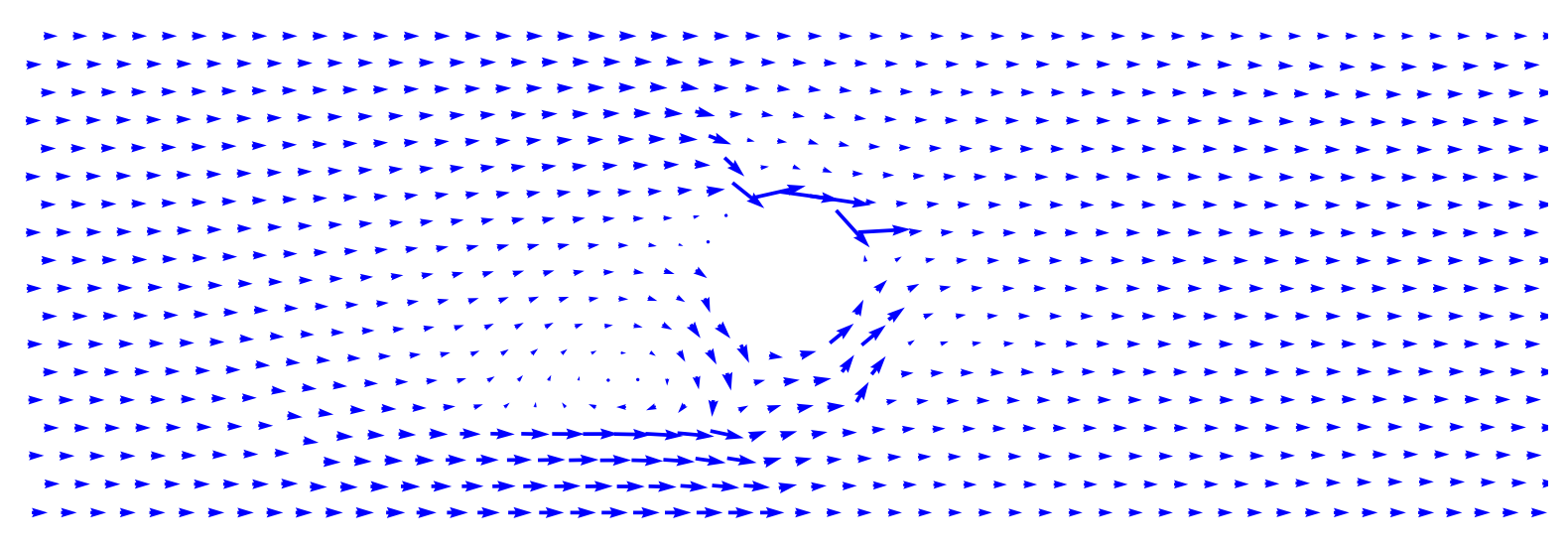}
			\label{fig:delta10b}
		\end{subfigure}
		\hfill
		\begin{subfigure}[t]{0.32\linewidth}
			\captionsetup{justification=raggedright, singlelinecheck=false, position=above}
			\caption{$\Delta {\bf r}$ for $\Delta t=1200$}
			\includegraphics[width=\linewidth]{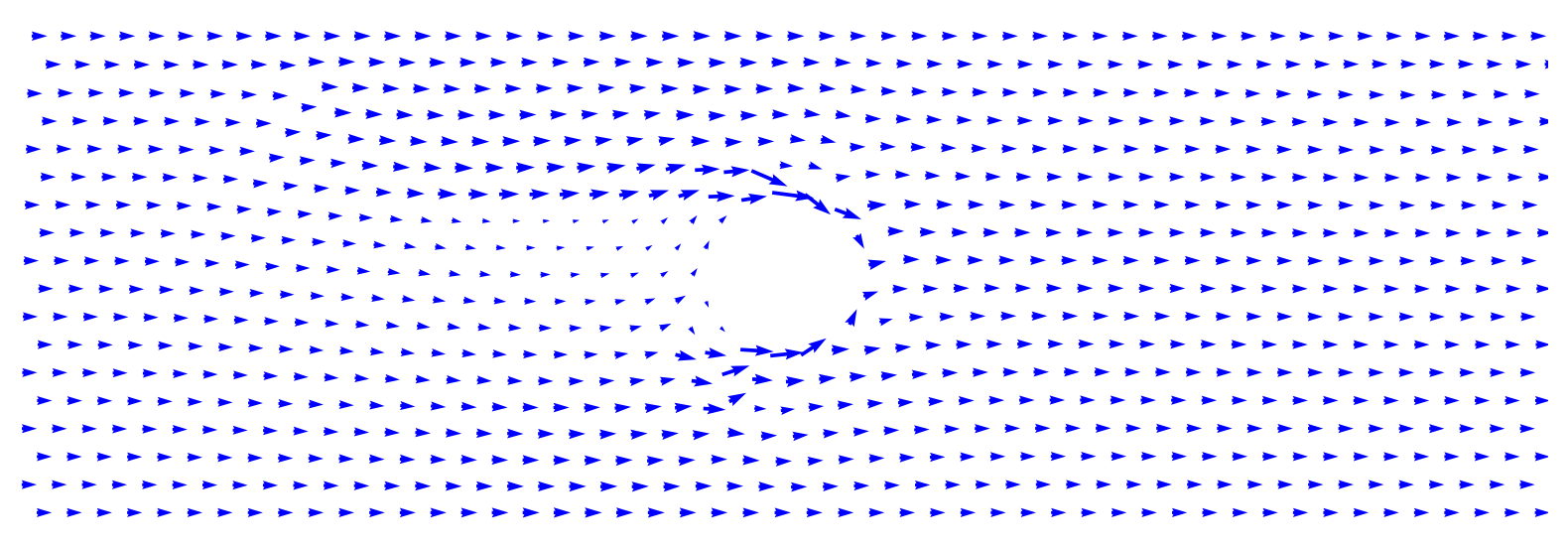}
			\label{fig:delta10c}
		\end{subfigure}
		\vspace{3mm}
		\begin{subfigure}[t]{0.32\linewidth}
			\captionsetup{justification=raggedright, singlelinecheck=false, position=above}
			\caption{$D^2_{\rm min}$ for $\Delta t=100$}
			\includegraphics[width=\linewidth]{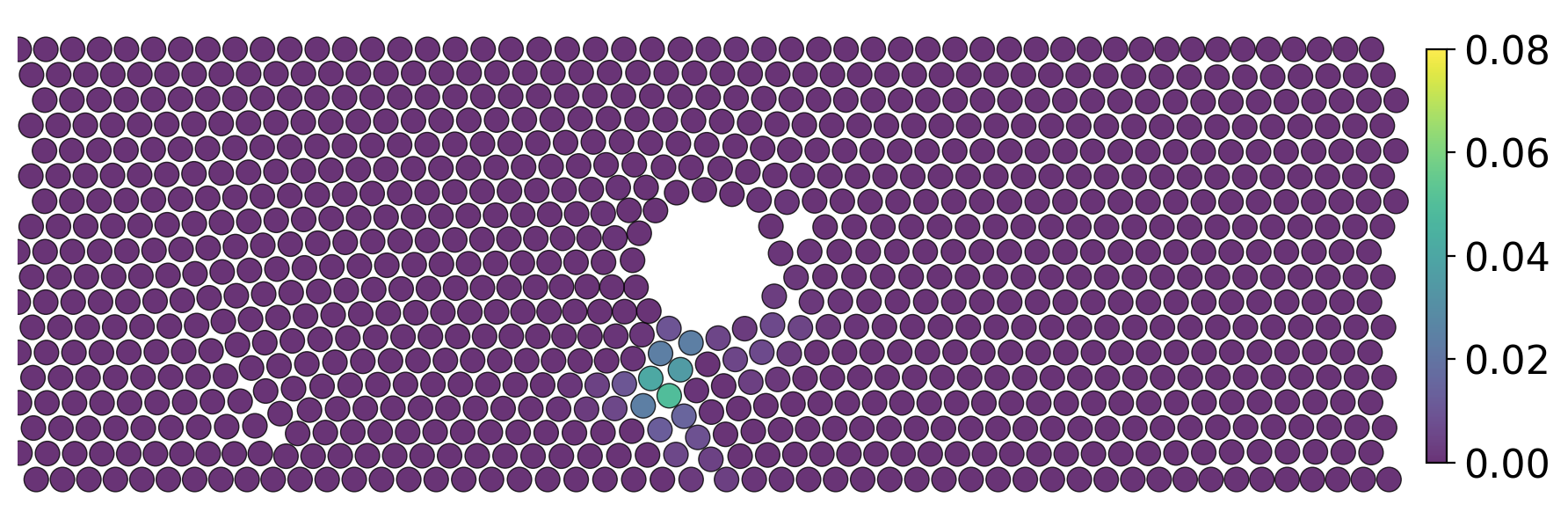}
			\label{fig:delta30a}
		\end{subfigure}
		\hfill
		\begin{subfigure}[t]{0.32\linewidth}
			\captionsetup{justification=raggedright, singlelinecheck=false, position=above}
			\caption{$D^2_{\rm min}$ for $\Delta t=300$}
			\includegraphics[width=\linewidth]{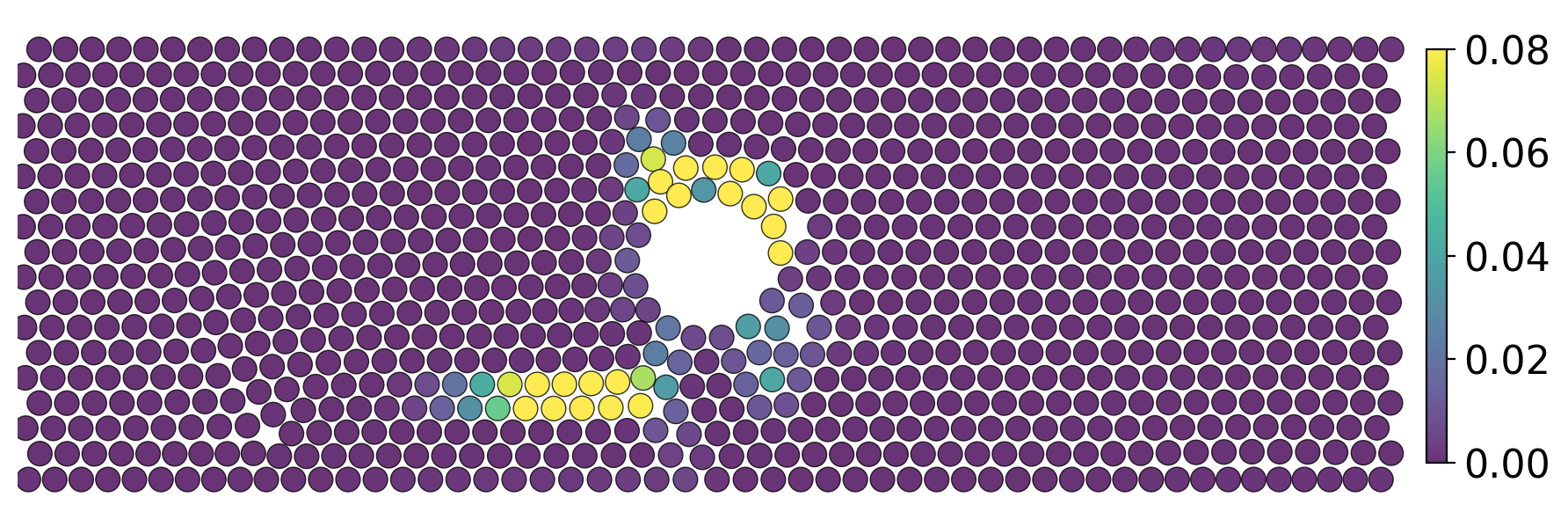}
			\label{fig:delta30b}
		\end{subfigure}
		\hfill
		\begin{subfigure}[t]{0.32\linewidth}
			\captionsetup{justification=raggedright, singlelinecheck=false, position=above}
			\caption{$D^2_{\rm min}$ for $\Delta t=1200$}
			\includegraphics[width=\linewidth]{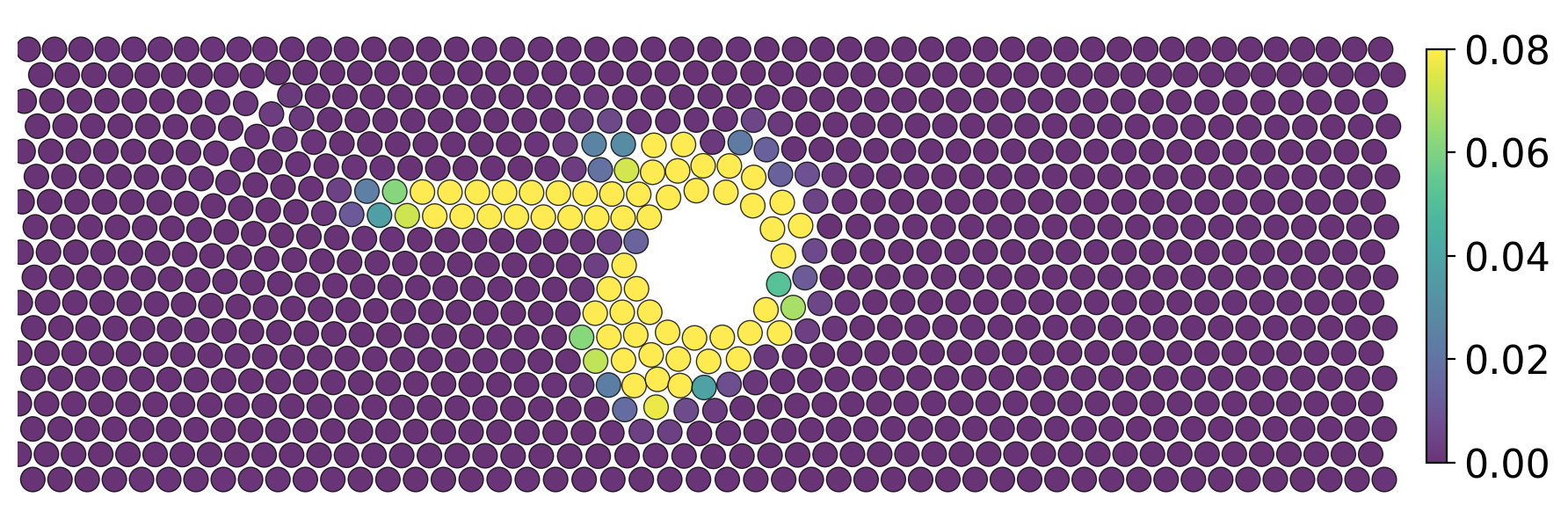}
			\label{fig:delta30c}
		\end{subfigure}
		\vspace{3mm}
		\begin{subfigure}[t]{0.32\linewidth}
			\captionsetup{justification=raggedright, singlelinecheck=false, position=above}
			\caption{Neighbor change events for $\Delta t=100$ }
			\includegraphics[width=\linewidth]{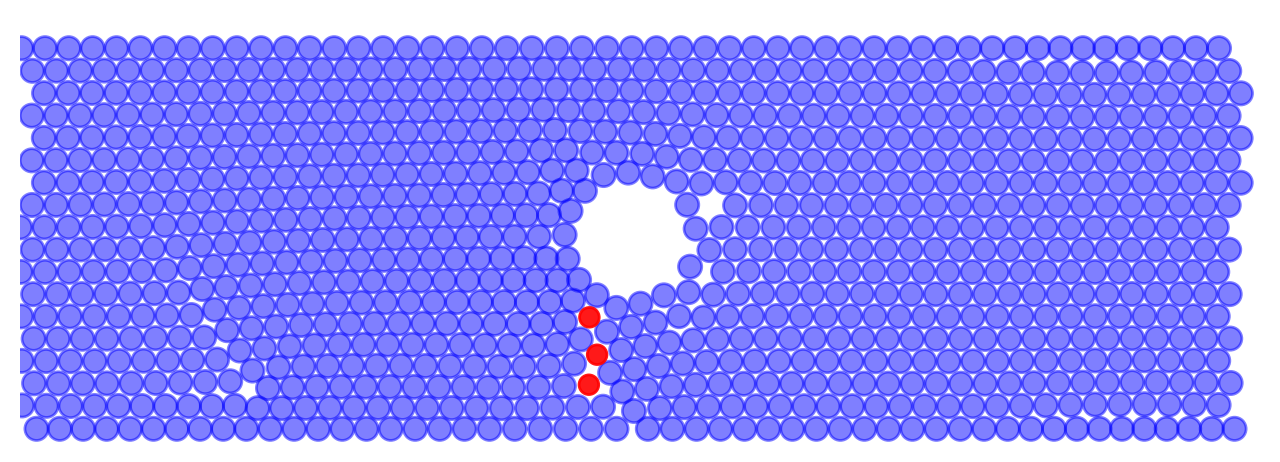}
			\label{fig:delta120a}
		\end{subfigure}
		\hfill
		\begin{subfigure}[t]{0.32\linewidth}
			\captionsetup{justification=raggedright, singlelinecheck=false, position=above}
			\caption{Neighbor change events for $\Delta t=300$}
			\includegraphics[width=\linewidth]{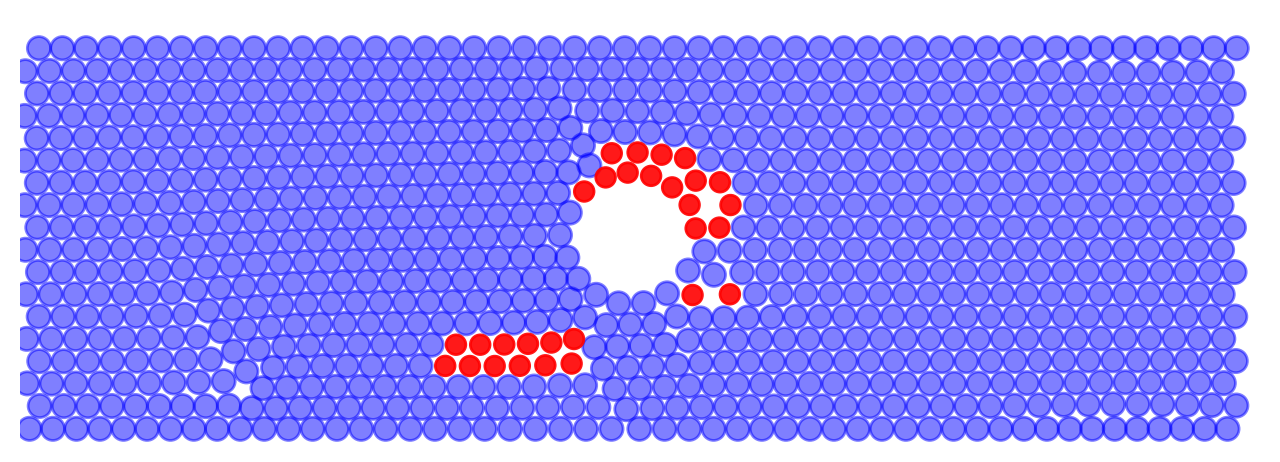}
			\label{fig:delta120b}
		\end{subfigure}
		\hfill
		\begin{subfigure}[t]{0.32\linewidth}
			\captionsetup{justification=raggedright, singlelinecheck=false, position=above}
			\caption{Neighbor change events for $\Delta t=1200$}
			\includegraphics[width=\linewidth]{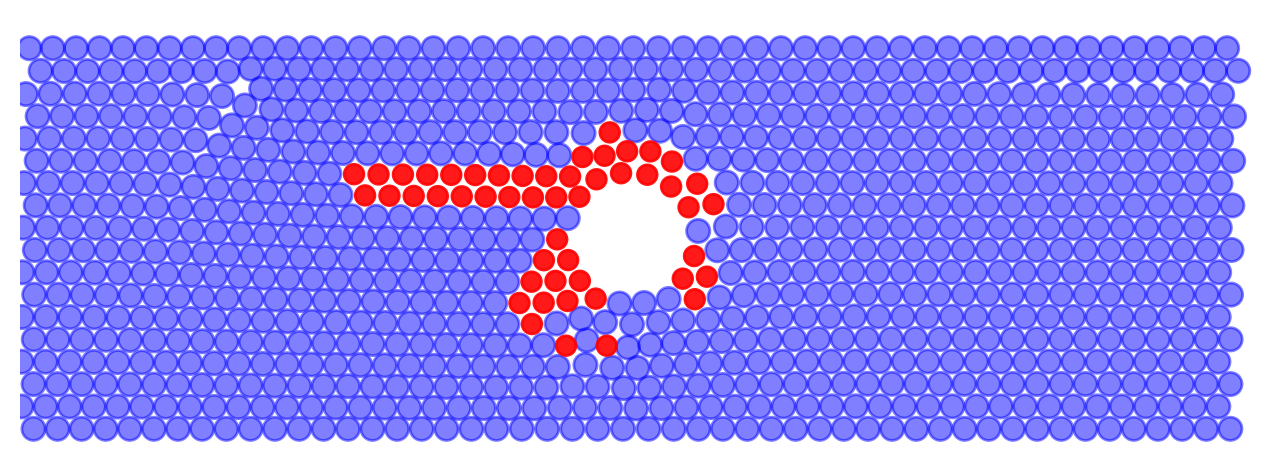}
			\label{fig:delta120c}
		\end{subfigure}
		\captionsetup{justification=raggedright,singlelinecheck=false}
		\caption{Representative flow behavior for $\delta = 0.001$.
			(a–c) Displacement vector field $\Delta \mathbf{r}_i$ computed over
			$\Delta t = 100 $ (a), 300 (b), and 1200 (c).
			The vector arrows are multiplied by 6, 1,
			and 0.5 for (a), (b), and (c),
			respectively. 
			(d–f) Minimum non-affine squared displacement $D_{\rm min}^2$ computed over
			$\Delta t = 100$ (d), 300 (e), and 1200 (f).
			The color bar indicates the magnitude of $D_{\rm min}^2$.
			(g–i) Neighbor change events detected over
			$\Delta t = 100$ (g), 300 (h), and 1200 (i).
			Red (blue) particles correspond to particles undergoing (not undergoing)
			a neighbor change event. }
		\label{fig:delta=0.001}
	\end{figure*}

	Figure~\ref{fig:delta=0.001} presents the case $\delta = 0.001$, which corresponds
	to a nearly monodisperse system.
	We monitor the displacement vector
	$\Delta \mathbf{r}_i = \mathbf{r}_i(t+\Delta t) - \mathbf{r}_i(t)$
	over a time interval $\Delta t$, which is systematically varied.
	Figures~\ref{fig:delta=0.001}(a), (b), and (c) correspond to
	$\Delta t = 100$ (short), 300 (intermediate), and 1200 (long) time scales,
	respectively.
	To improve visibility, the displacement vectors are rescaled by constant
	factors (see figure caption).
	
	One observes large displacements localized near the obstacle at short time scales,
	while most of the system exhibits relatively smooth motion.
	At intermediate and longer time scales, sliding-like motion connected to the
	central obstacle develops in a complex spatial pattern.
	These results demonstrate that the central obstacle induces complex and heterogeneous plastic behavior even in nearly monodisperse systems. In contrast, without the central obstacle, no plastic phenomena take place under uniform channel flow.

	To highlight plastic activity and remove the affine contribution associated with
	the net flow, we compute the minimum non-affine squared displacement
	$D_{\rm min}^2$ (see Appendix~\ref{sec:D2min} for the definition).
	Results are shown in Figs.~\ref{fig:delta=0.001}(d–f), corresponding to the same
	time intervals as in panels (a–c).
	The plots indeed show that plastic events take place near the obstacle and
	propagate in a highly directional, sliding-like manner.
	
	To further support these observations, we introduce a binary indicator of plastic rearrangements, which takes the value 1 (red) if a neighbor change event occurs and 0 (blue) otherwise. The precise definition of a neighbor change event is given in Appendix~\ref{sec:T1}. This includes the conventional T1 event, corresponding to a neighbor swapping among four adjacent particles, as a special case. We show maps of the neighbor change event indicator in Figs.~\ref{fig:delta=0.001}(g–i), which correspond to the same trajectories and time intervals as those shown in panels (a–c) and (d–f). This binarized representation consistently confirms the preceding observations.
	We observe sliding-like motion along the $x$-direction (the direction of the external drive) in Fig.~\ref{fig:delta=0.001}, yet we also observe sliding motion along the $x$–$y$ diagonal directions in some other samples. Most of the samples we studied show plastic events induced near the obstacle; however, occasionally, plastic events are triggered far from the obstacle due to the long-range nature of elastic interactions. In Fig.~\ref{fig:sliding} of Appendix~\ref{sec:sliding}, we present additional examples illustrating such sliding motions, as well as a sample showing a sliding plastic event occurring away from the obstacle.

	\subsection{High polydispersity: Amorphous system}

	\begin{figure*}[htbp]
		\centering
		\begin{subfigure}[t]{0.32\linewidth}
			\captionsetup{justification=raggedright, singlelinecheck=false, position=above}  
			\caption{$\Delta {\bf r}$ for $\Delta t=100$}
			\includegraphics[width=\linewidth]{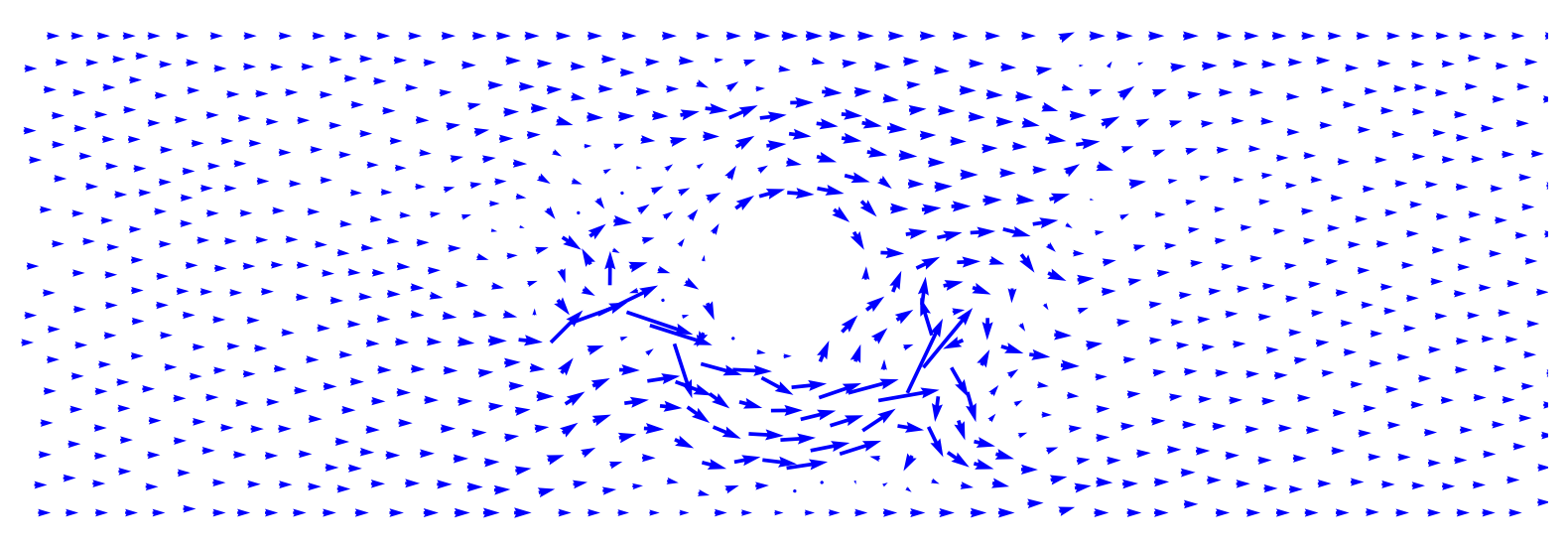}
			\label{fig:delta10a}
		\end{subfigure}
		\hfill
		\begin{subfigure}[t]{0.32\linewidth}
			\captionsetup{justification=raggedright, singlelinecheck=false, position=above}
			\caption{$\Delta {\bf r}$ for $\Delta t=300$}
			\includegraphics[width=\linewidth]{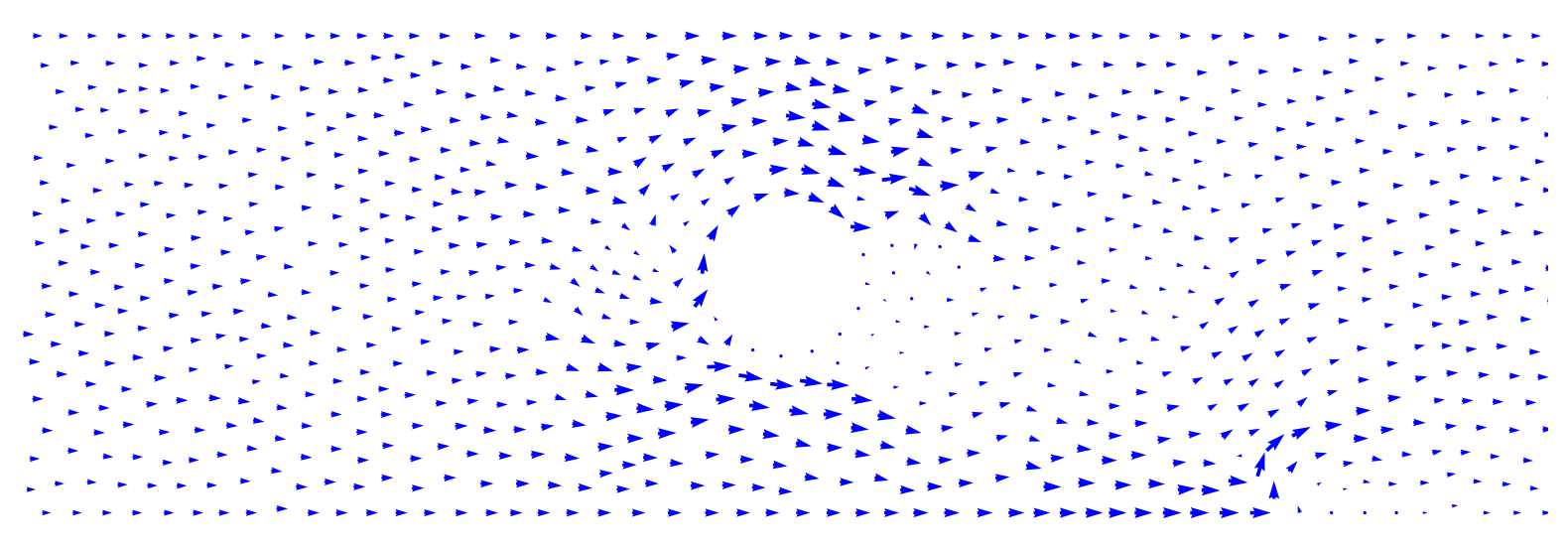}
			\label{fig:delta10b}
		\end{subfigure}
		\hfill
		\begin{subfigure}[t]{0.32\linewidth}
			\captionsetup{justification=raggedright, singlelinecheck=false, position=above}
			\caption{$\Delta {\bf r}$ for $\Delta t=1200$}
			\includegraphics[width=\linewidth]{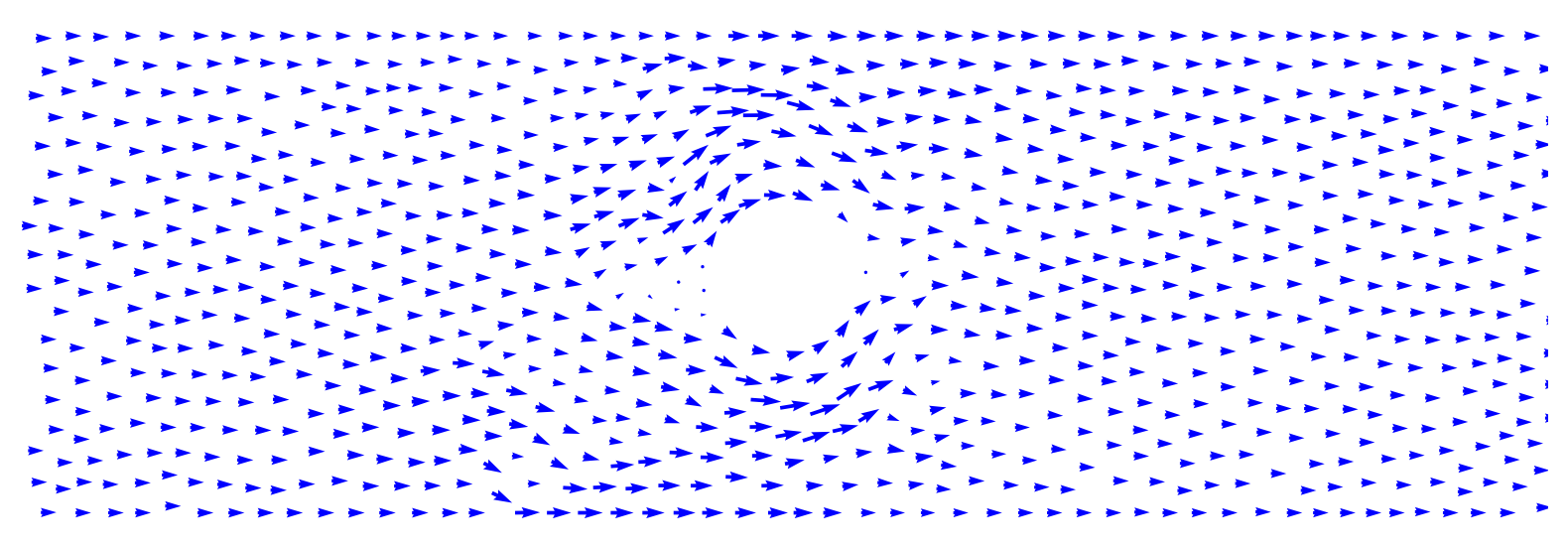}
			\label{fig:delta10c}
		\end{subfigure}
		\vspace{3mm}
		\begin{subfigure}[t]{0.32\linewidth}
			\captionsetup{justification=raggedright, singlelinecheck=false, position=above}
			\caption{$D^2_{\rm min}$ for $\Delta t=100$}
			\includegraphics[width=\linewidth]{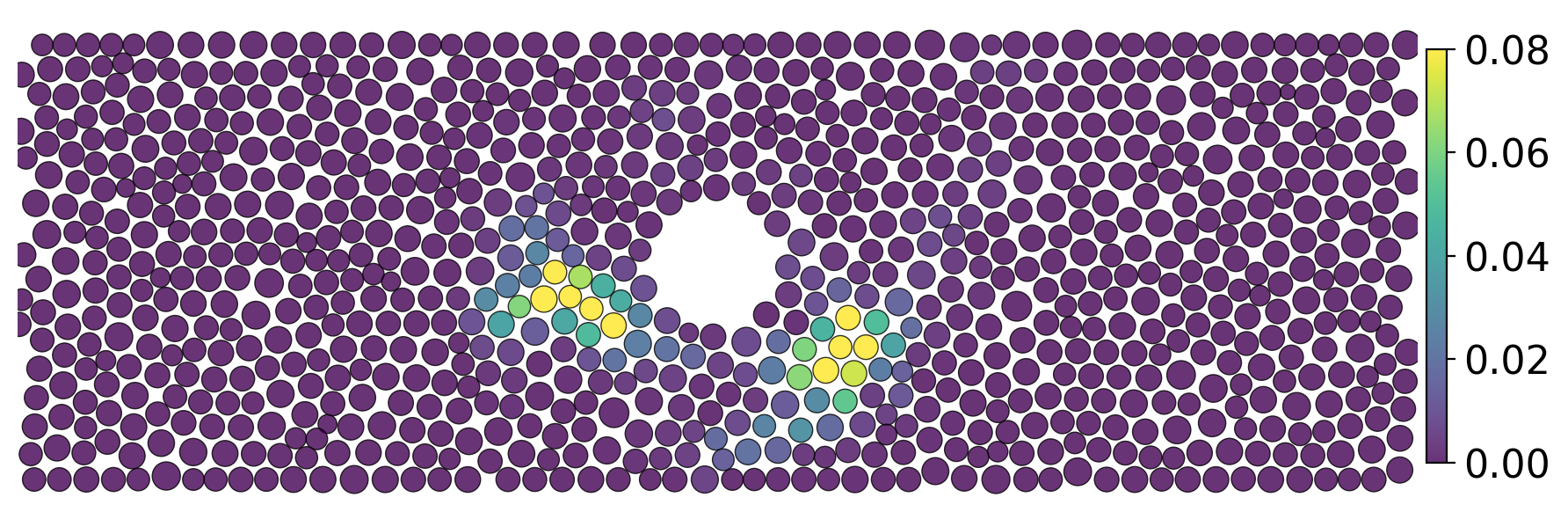}
			\label{fig:delta30a}
		\end{subfigure}
		\hfill
		\begin{subfigure}[t]{0.32\linewidth}
			\captionsetup{justification=raggedright, singlelinecheck=false, position=above}
			\caption{$D^2_{\rm min}$ for $\Delta t=300$}
			\includegraphics[width=\linewidth]{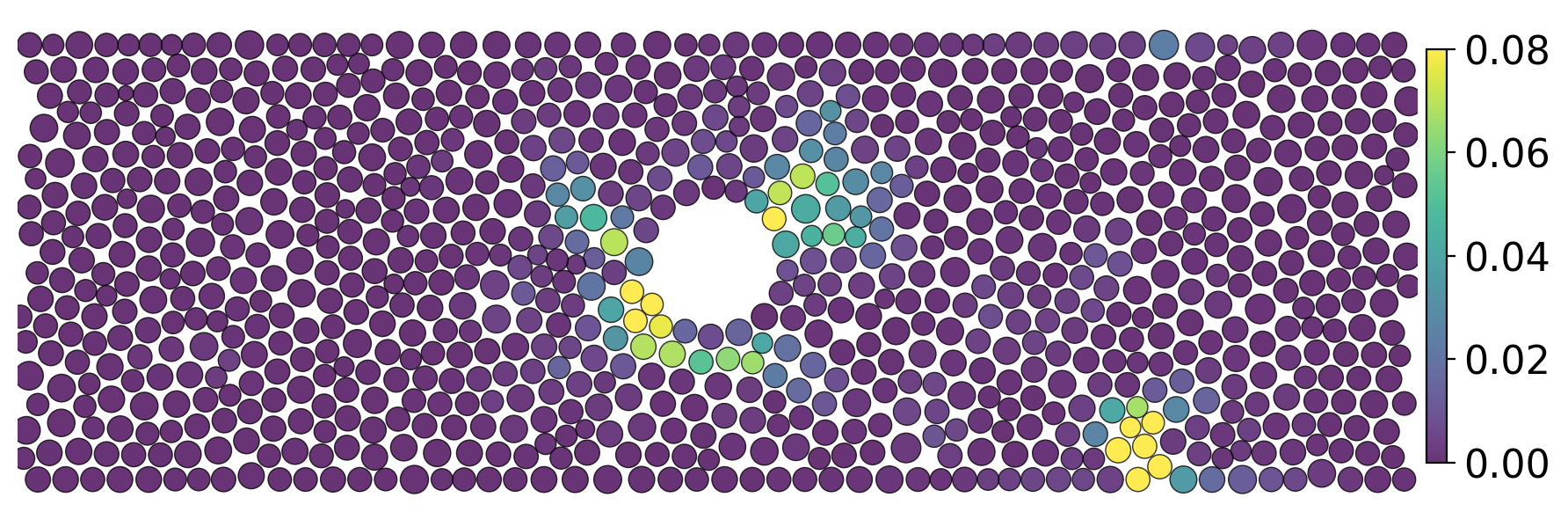}
			\label{fig:delta30b}
		\end{subfigure}
		\hfill
		\begin{subfigure}[t]{0.32\linewidth}
			\captionsetup{justification=raggedright, singlelinecheck=false, position=above}
			\caption{$D^2_{\rm min}$ for $\Delta t=1200$}
			\includegraphics[width=\linewidth]{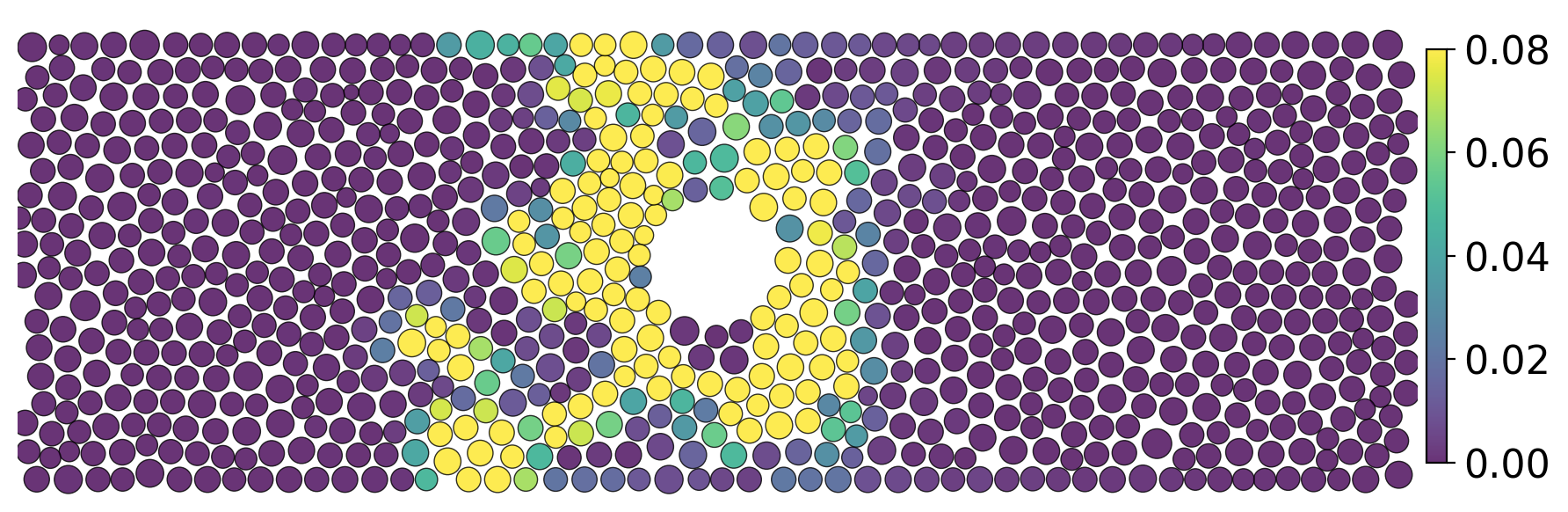}
			\label{fig:delta30c}
		\end{subfigure}
		\vspace{3mm}
		\begin{subfigure}[t]{0.32\linewidth}
			\captionsetup{justification=raggedright, singlelinecheck=false, position=above}
			\caption{Neighbor change events for $\Delta t=100$}
			\includegraphics[width=\linewidth]{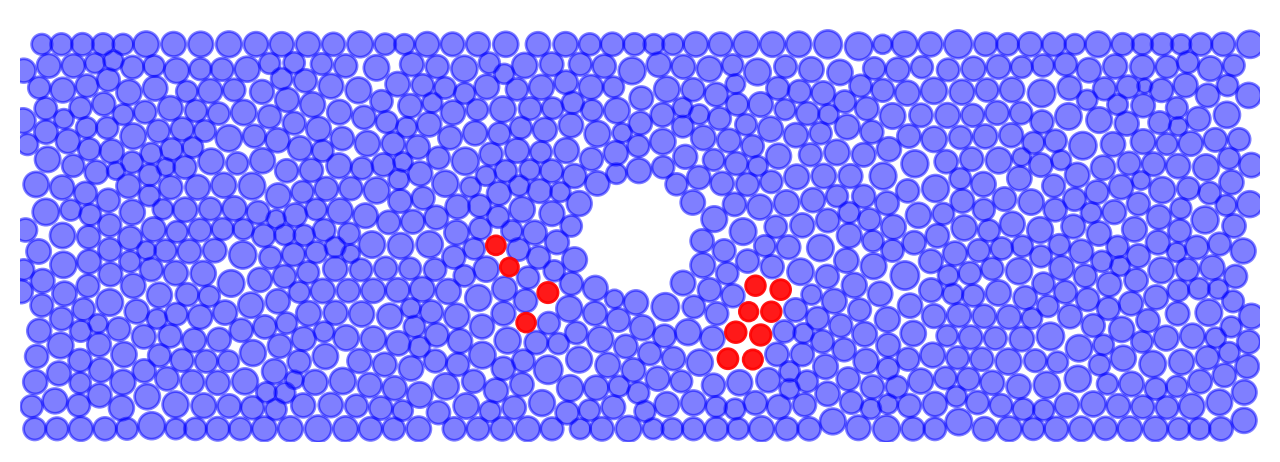}
			\label{fig:delta120a}
		\end{subfigure}
		\hfill
		\begin{subfigure}[t]{0.32\linewidth}
			\captionsetup{justification=raggedright, singlelinecheck=false, position=above}
			\caption{Neighbor change events for $\Delta t=300$}
			\includegraphics[width=\linewidth]{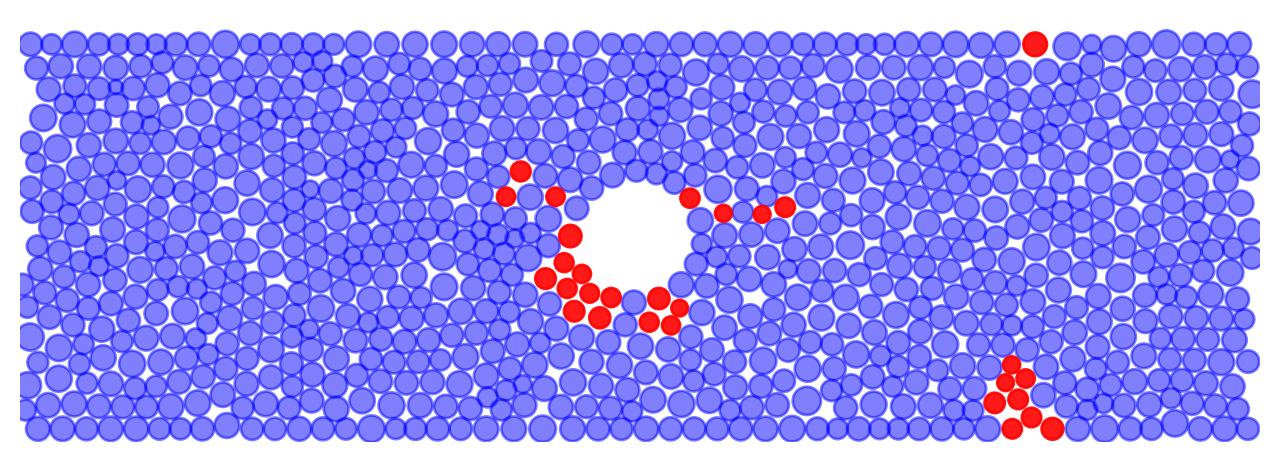}
			\label{fig:delta120b}
		\end{subfigure}
		\hfill
		\begin{subfigure}[t]{0.32\linewidth}
			\captionsetup{justification=raggedright, singlelinecheck=false, position=above}
			\caption{Neighbor change events for $\Delta t=1200$}
			\includegraphics[width=\linewidth]{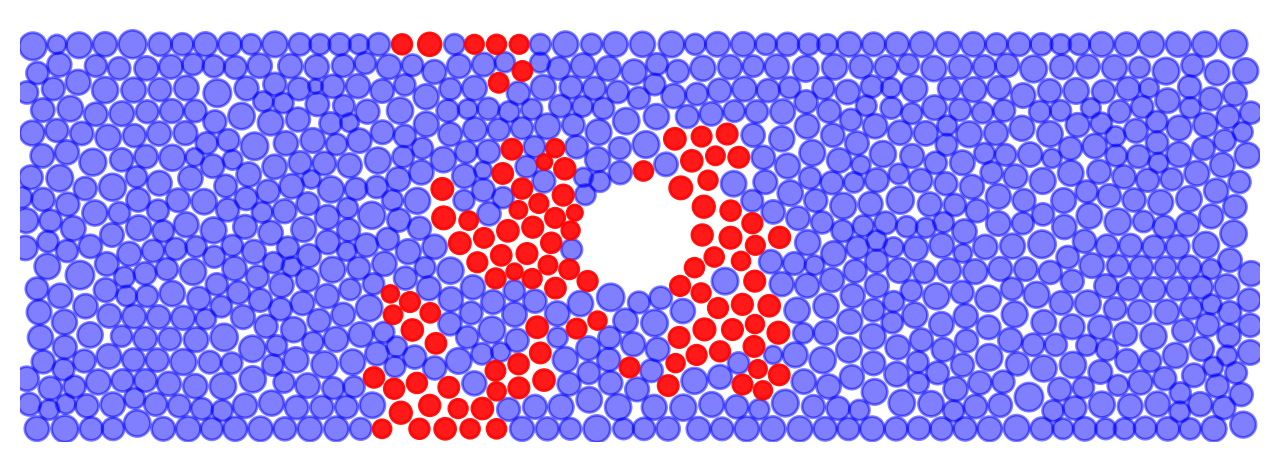}
			\label{fig:delta120c}
		\end{subfigure}
		\captionsetup{justification=raggedright,singlelinecheck=false}
		\caption{Same as Fig.~\ref{fig:delta=0.001} but for $\delta = 0.15$.
			The displacement vectors $\Delta \mathbf{r}_i$ are multiplied by 5, 1.5, and 0.5
			for (a), (b), and (c), respectively.}
		\label{fig:delta=0.15}
	\end{figure*}
	
	We next turn our attention to the highly polydisperse system with
	$\delta = 0.15$, shown in Fig.~\ref{fig:delta=0.15}.
	As shown in Fig.~\ref{fig:delta=0.15}(a), at short time scales we again observe
	large displacements near the obstacle.
	However, the displacement directions are highly scattered, reminiscent of
	bulk amorphous materials under shear~\cite{maloney2006amorphous}.
	This behavior is further characterized by localized plastic events,
	clearly visible in both $D_{\rm min}^2$ and the neighbor change event maps in
	Figs.~\ref{fig:delta=0.15}(d) and (g), respectively.
	
	Interestingly, as time increases, plastic activity remains largely isotropic,
	in contrast to the directional, sliding-like motion observed in the
	crystalline system with $\delta = 0.001$ in Fig.~\ref{fig:delta=0.001}.
	This difference can be attributed to the effect of polydispersity, which
	introduces structural disorder and suppresses coherent directional motion,
	favoring instead more random and isotropic rearrangements.

	\subsection{Intermediate polydispersity}
	
	\begin{figure*}[htbp]
		\centering
		\begin{subfigure}[t]{0.32\linewidth}
			\captionsetup{justification=raggedright, singlelinecheck=false, position=above}  
			\caption{$\Delta {\bf r}$ for $\Delta t=100$}
			\includegraphics[width=\linewidth]{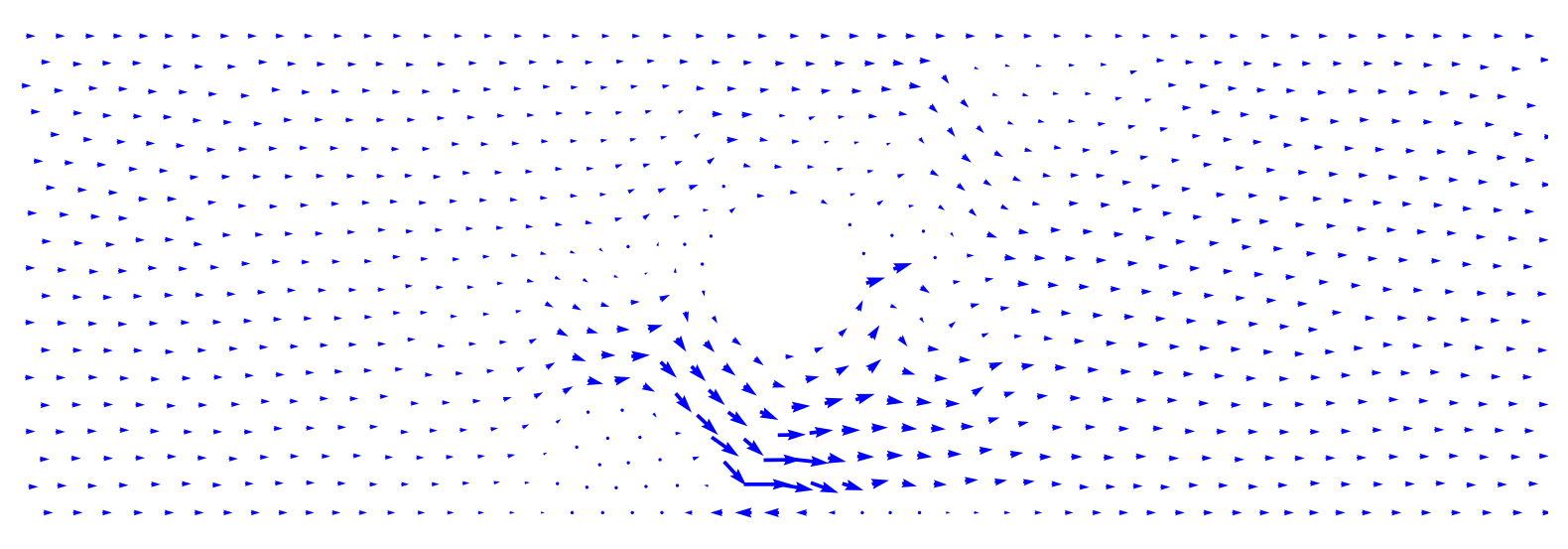}
			\label{fig:delta10a}
		\end{subfigure}
		\hfill
		\begin{subfigure}[t]{0.32\linewidth}
			\captionsetup{justification=raggedright, singlelinecheck=false, position=above}
			\caption{$\Delta {\bf r}$ for $\Delta t=300$}
			\includegraphics[width=\linewidth]{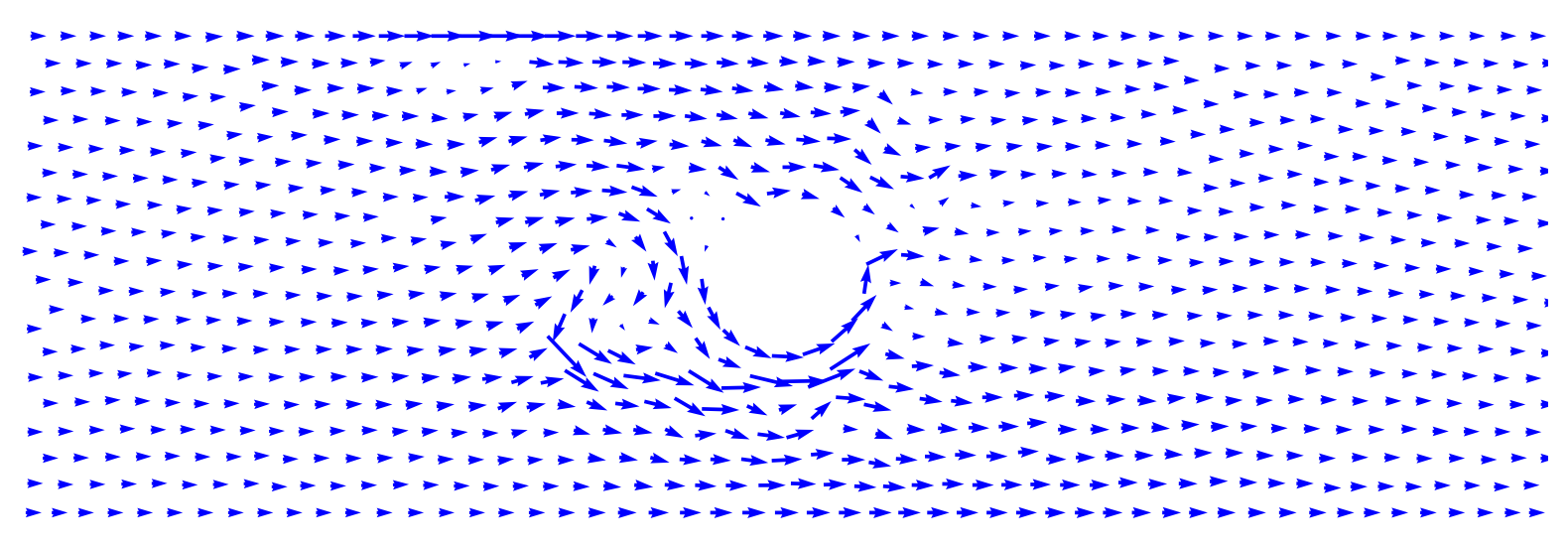}
			\label{fig:delta10b}
		\end{subfigure}
		\hfill
		\begin{subfigure}[t]{0.32\linewidth}
			\captionsetup{justification=raggedright, singlelinecheck=false, position=above}
			\caption{$\Delta {\bf r}$ for $\Delta t=1200$}
			\includegraphics[width=\linewidth]{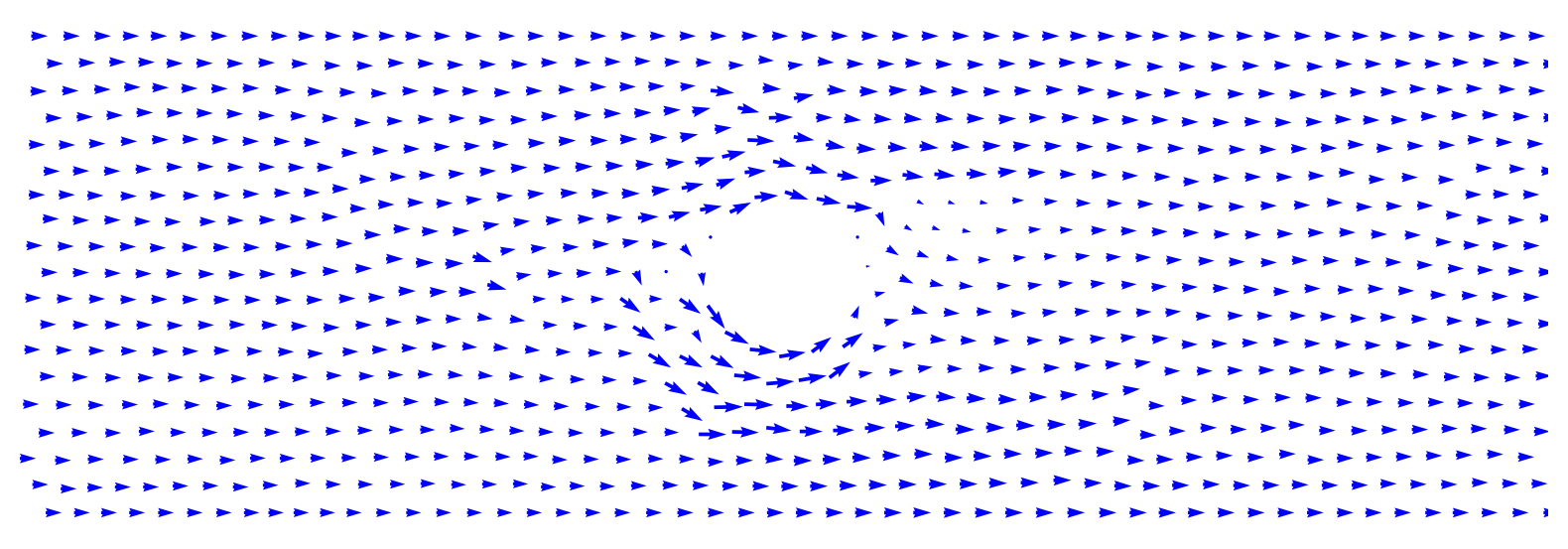}
			\label{fig:delta10c}
		\end{subfigure}
		\vspace{3mm}
		\begin{subfigure}[t]{0.32\linewidth}
			\captionsetup{justification=raggedright, singlelinecheck=false, position=above}
			\caption{$D^2_{\rm min}$ for $\Delta t=100$}
			\includegraphics[width=\linewidth]{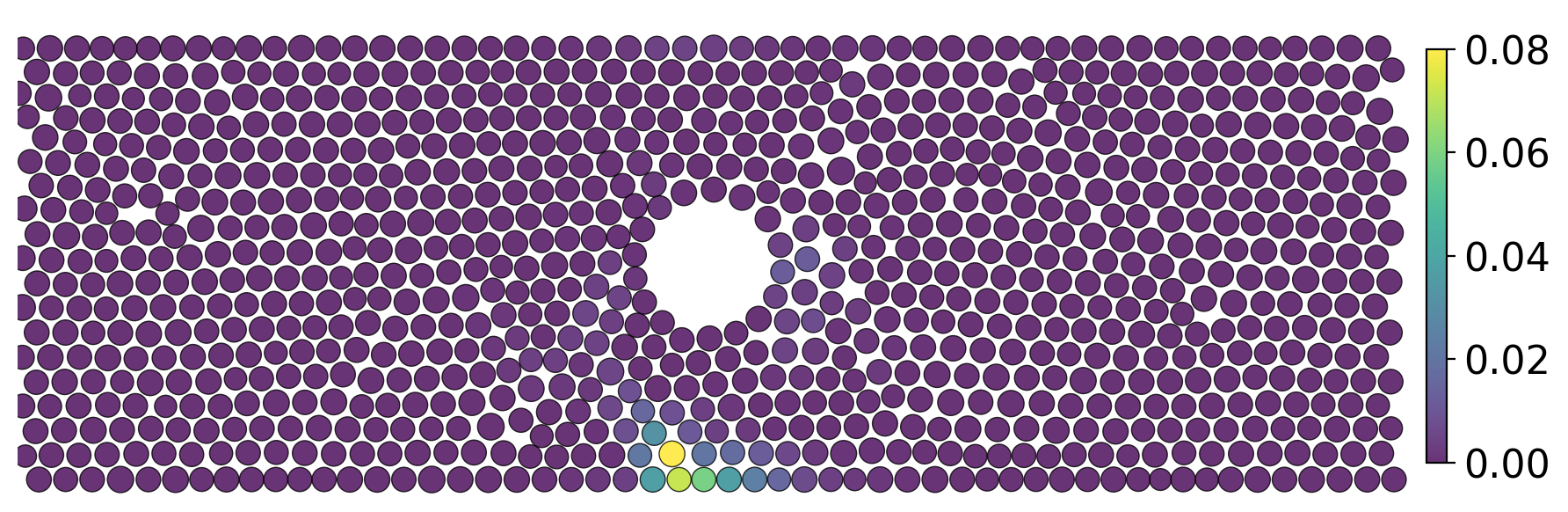}
			\label{fig:delta30a}
		\end{subfigure}
		\hfill
		\begin{subfigure}[t]{0.32\linewidth}
			\captionsetup{justification=raggedright, singlelinecheck=false, position=above}
			\caption{$D^2_{\rm min}$ for $\Delta t=300$}
			\includegraphics[width=\linewidth]{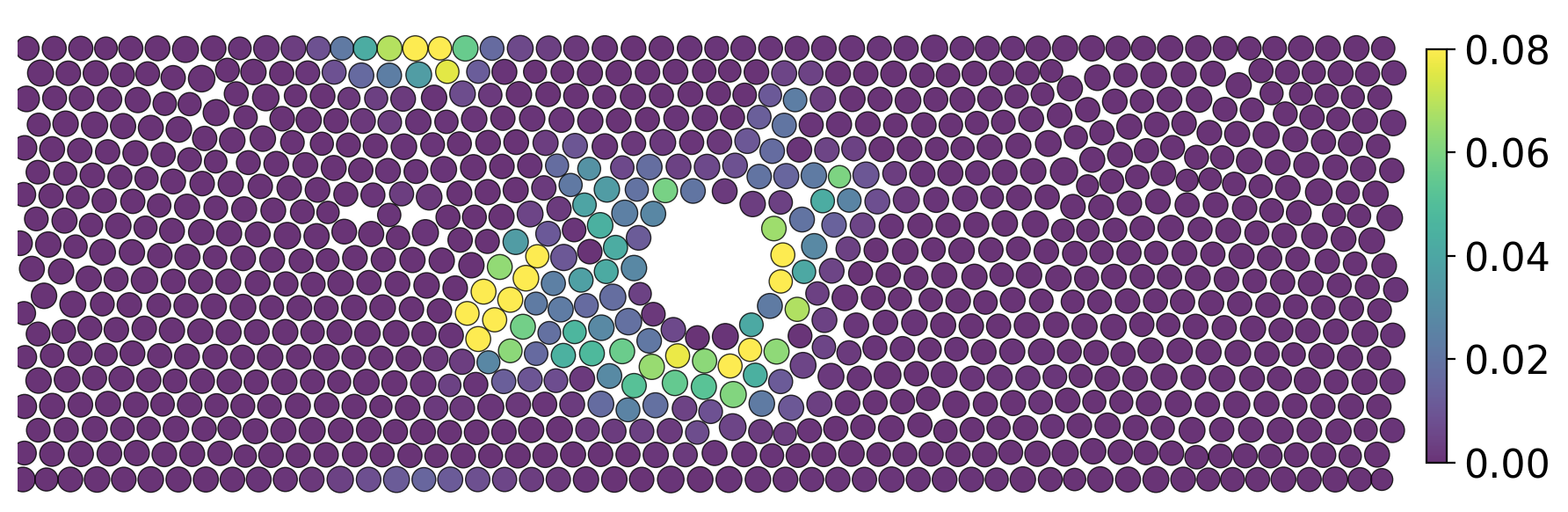}
			\label{fig:delta30b}
		\end{subfigure}
		\hfill
		\begin{subfigure}[t]{0.32\linewidth}
			\captionsetup{justification=raggedright, singlelinecheck=false, position=above}
			\caption{$D^2_{\rm min}$ for $\Delta t=1200$}
			\includegraphics[width=\linewidth]{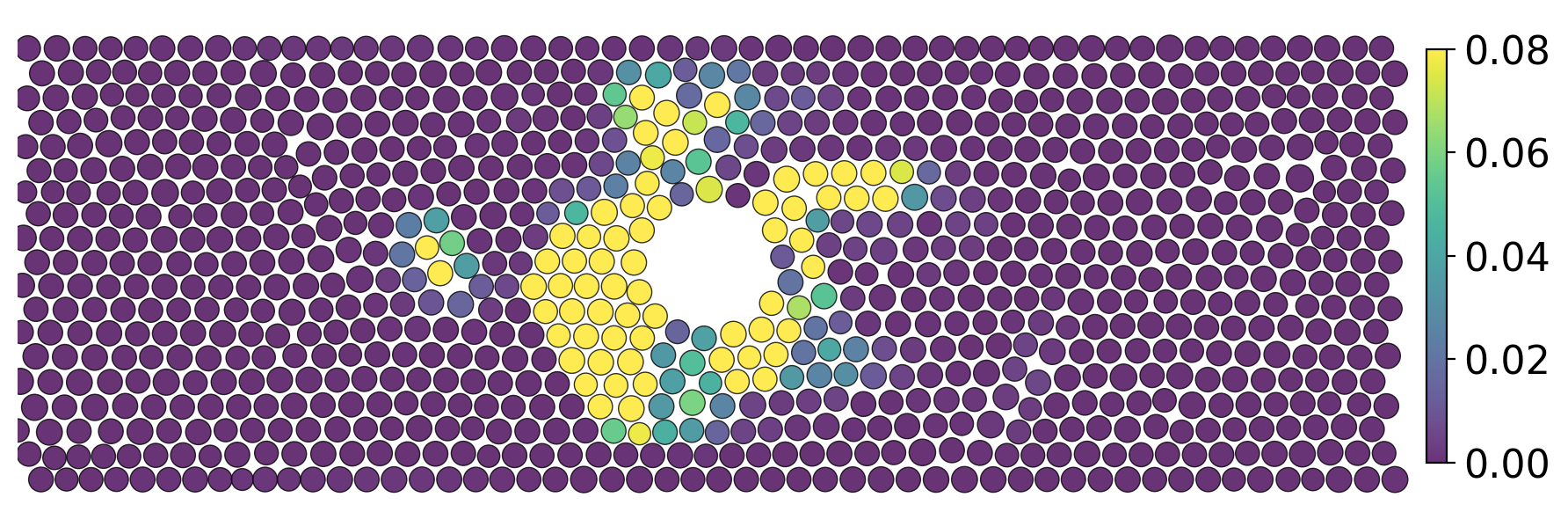}
			\label{fig:delta30c}
		\end{subfigure}
		\vspace{3mm}
		\begin{subfigure}[t]{0.32\linewidth}
			\captionsetup{justification=raggedright, singlelinecheck=false, position=above}
			\caption{Neighbor change events for $\Delta t=100$}
			\includegraphics[width=\linewidth]{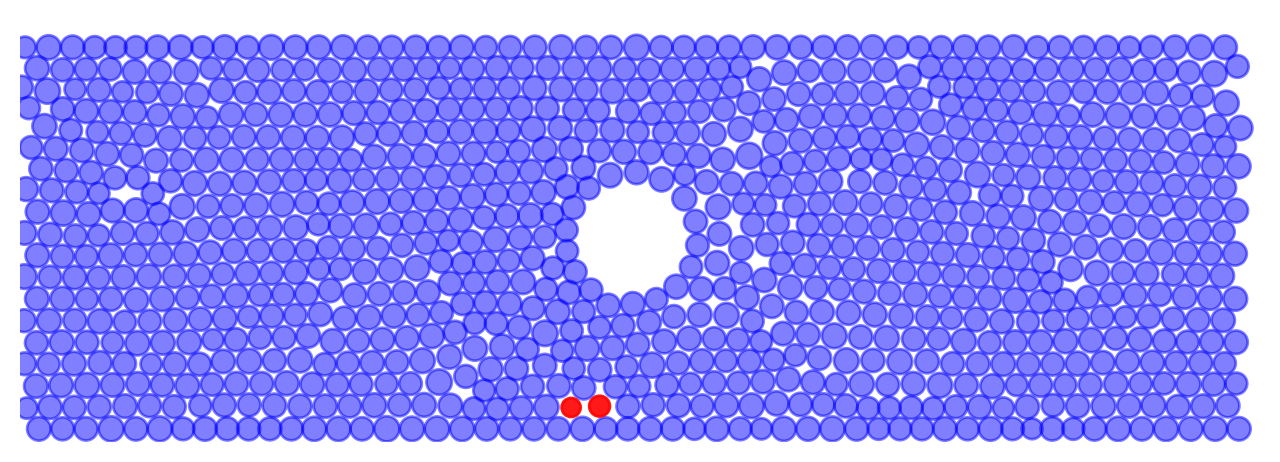}
			\label{fig:delta120a}
		\end{subfigure}
		\hfill
		\begin{subfigure}[t]{0.32\linewidth}
			\captionsetup{justification=raggedright, singlelinecheck=false, position=above}
			\caption{Neighbor change events for $\Delta t=300$}
			\includegraphics[width=\linewidth]{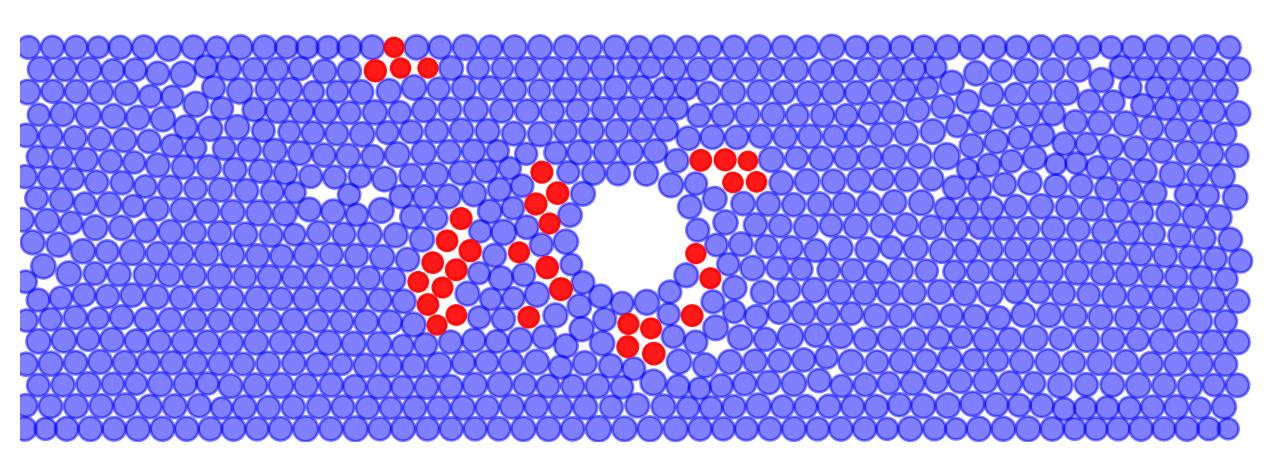}
			\label{fig:delta120b}
		\end{subfigure}
		\hfill
		\begin{subfigure}[t]{0.32\linewidth}
			\captionsetup{justification=raggedright, singlelinecheck=false, position=above}
			\caption{Neighbor change events for $\Delta t=1200$}
			\includegraphics[width=\linewidth]{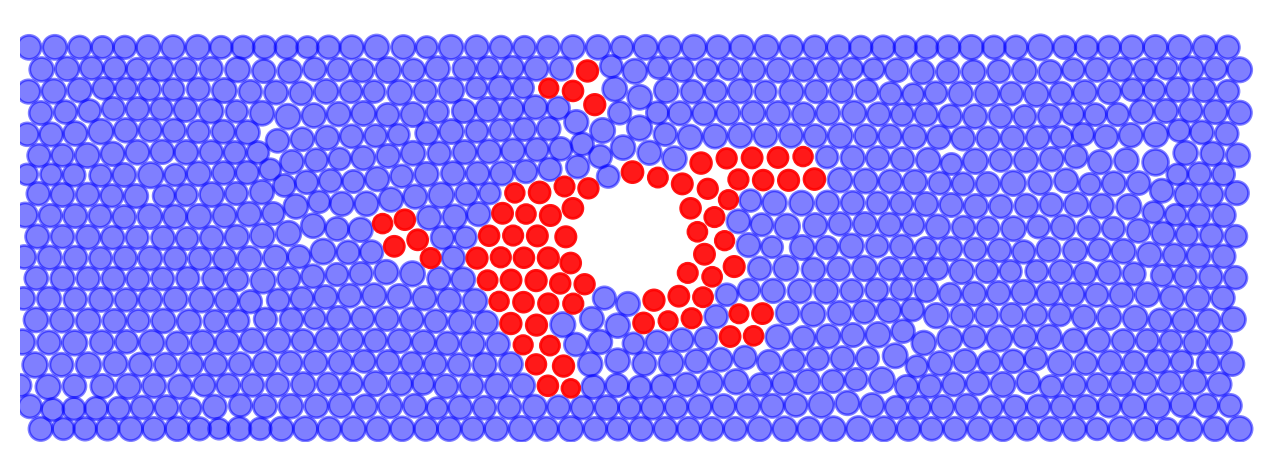}
			\label{fig:delta120c}
		\end{subfigure}
		\captionsetup{justification=raggedright,singlelinecheck=false}
		\caption{Same as Fig.~\ref{fig:delta=0.001} but for $\delta = 0.07$.
			The displacement vectors $\Delta \mathbf{r}_i$ are multiplied by 4, 2, and 0.5
			for (a), (b), and (c), respectively.}
		\label{fig:delta=0.07}
	\end{figure*}
	
	We also examine the system with intermediate polydispersity, $\delta = 0.07$, shown in Fig.~\ref{fig:delta=0.07}. Overall, we observe mixed features characteristic of both the crystalline-like and amorphous regimes. While some degree of directional, sliding-like motion persists near the obstacle, plastic activity also displays more localized and isotropic rearrangements, indicative of increasing structural disorder.

	\subsection{Anisotropy of plastic rearrangements}
	
	We have observed that polydispersity strongly affects plastic behavior,
	in particular the spatial structure of rearrangements.
	While highly polydisperse systems exhibit heterogeneous plasticity that is
	predominantly isotropic, low-polydispersity, crystalline-like systems display
	highly directional, sliding-like motion. The latter behavior is strongly reminiscent of dislocation gliding, responsible for the plasticity of crystalline solids~\cite{sethna2017deformation,ghimenti2024shear}.
	
	To quantify this crossover, we introduce an anisotropy parameter $A$, which distinguishes directional, sliding-like rearrangements in weakly polydisperse crystalline systems from isotropic, localized rearrangements in highly polydisperse amorphous systems. Below, we describe how $A$ is constructed.

\begin{figure*}[htbp]
		\centering
		\scalebox{0.8}{
			\begin{tabular}{@{}c@{\hspace{5cm}}c@{}}
				\begin{subfigure}[t]{0.48\linewidth}
					\captionsetup{justification=raggedright,singlelinecheck=false,position=above}
					\caption{$\delta=0.001$: $\Delta {\bf r}$ for $\Delta t=900$}
					\includegraphics[width=\linewidth]{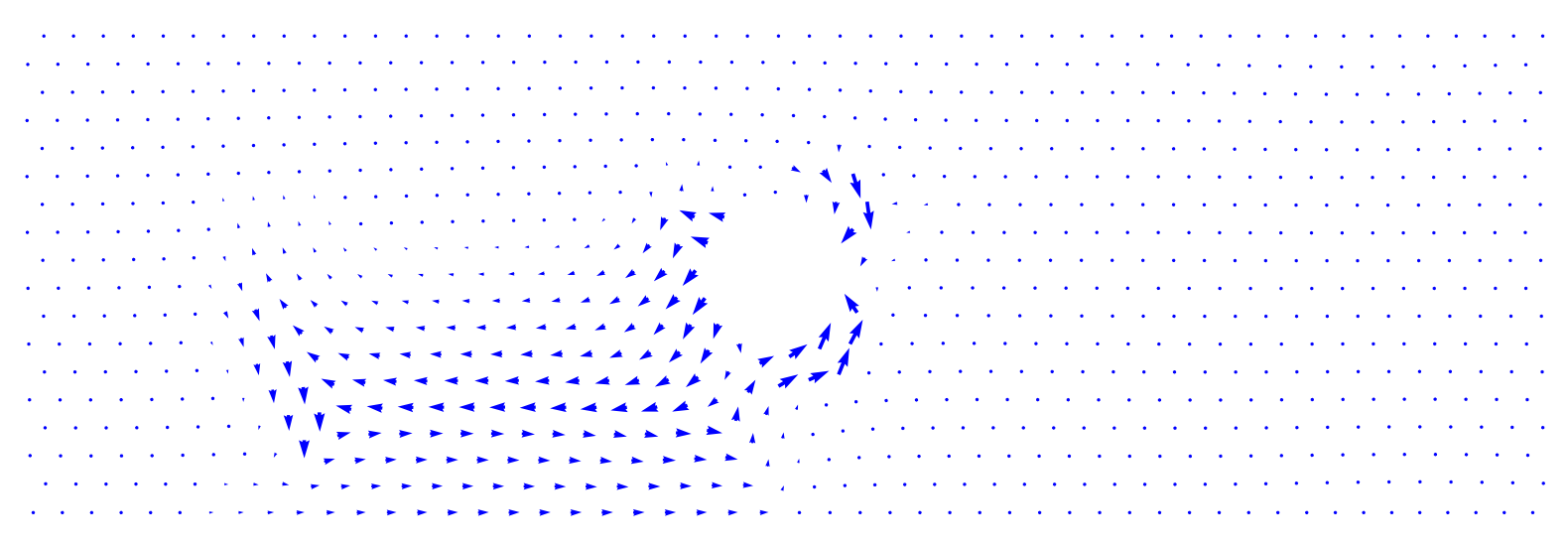}
				\end{subfigure}
				&
				\begin{subfigure}[t]{0.48\linewidth}
					\captionsetup{justification=raggedright,singlelinecheck=false,position=above}
					\caption{$\delta=0.15$: $\Delta {\bf r}$ for $\Delta t=900$}
					\includegraphics[width=\linewidth]{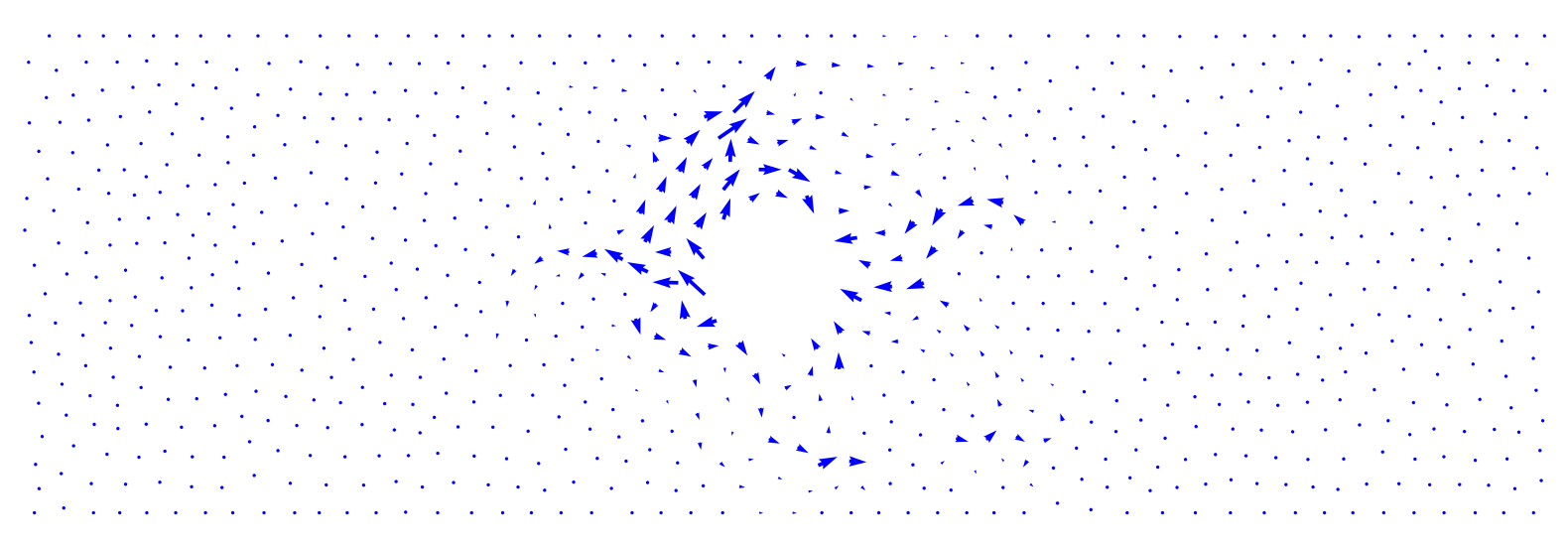}
				\end{subfigure}
				\\[4pt]
				
				\begin{subfigure}[t]{0.48\linewidth}
					\captionsetup{justification=raggedright,singlelinecheck=false,position=above}
					\caption{$\delta=0.001$: $D^2_{\rm min}$ for $\Delta t=900$}
					\includegraphics[width=\linewidth]{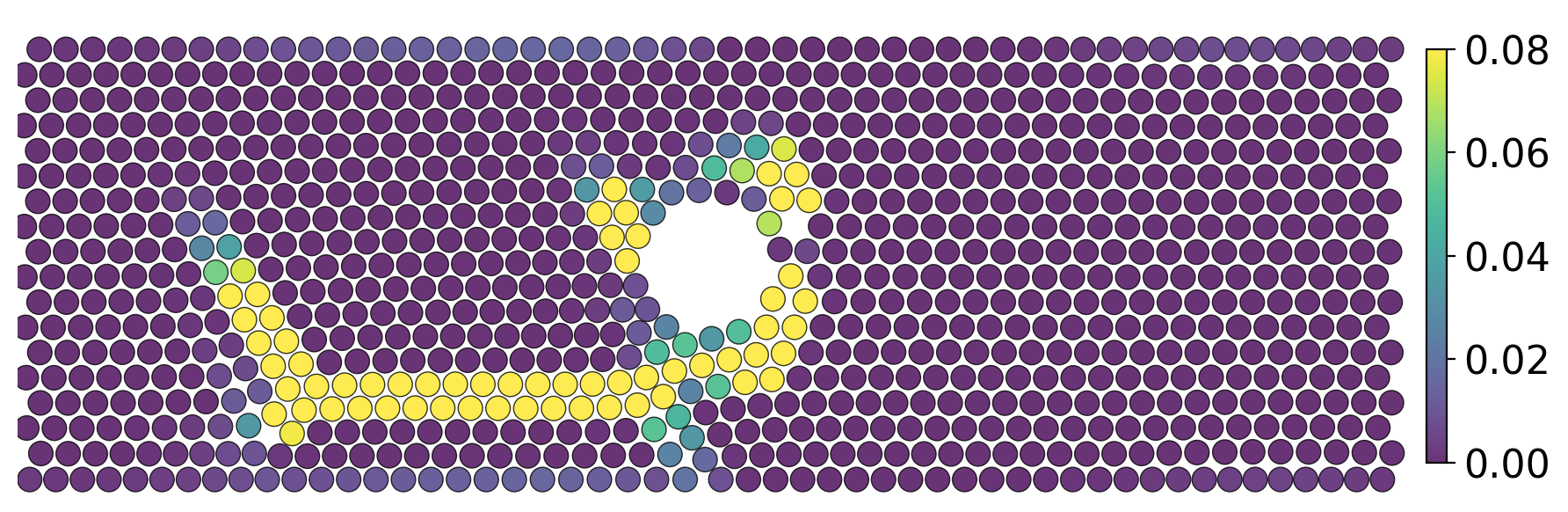}
				\end{subfigure}
				&
				\begin{subfigure}[t]{0.48\linewidth}
					\captionsetup{justification=raggedright,singlelinecheck=false,position=above}
					\caption{$\delta=0.15$: $D^2_{\rm min}$ for $\Delta t=900$}
					\includegraphics[width=\linewidth]{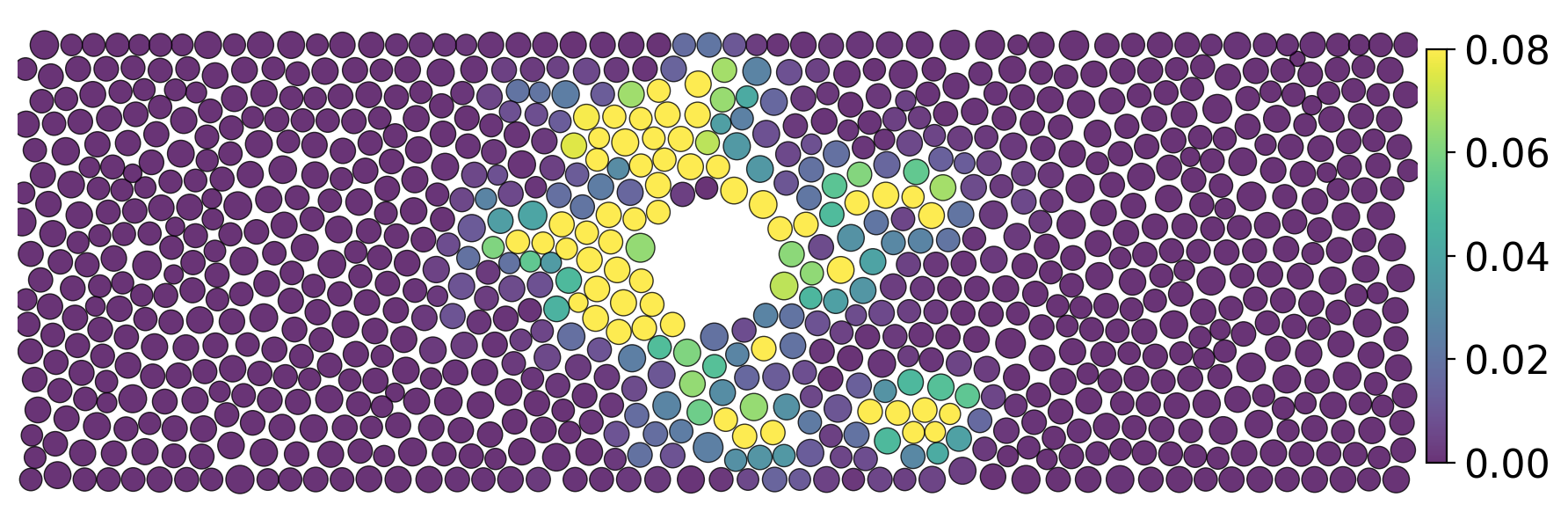}
				\end{subfigure}
				\\[4pt]
				
				\begin{subfigure}[t]{0.48\linewidth}
					\captionsetup{justification=raggedright,singlelinecheck=false,position=above}
					\caption{$\delta=0.001$: Neighbor change events for $\Delta t=900$}
					\includegraphics[width=\linewidth]{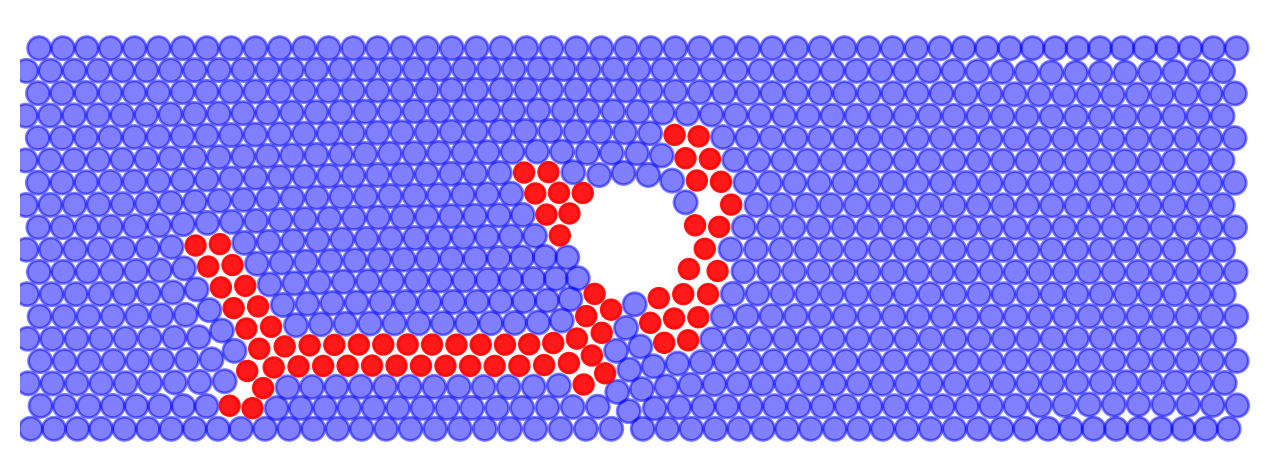}
				\end{subfigure}
				&
				\begin{subfigure}[t]{0.48\linewidth}
					\captionsetup{justification=raggedright,singlelinecheck=false,position=above}
					\caption{$\delta=0.15$: Neighbor change events for $\Delta t=900$}
					\includegraphics[width=\linewidth]{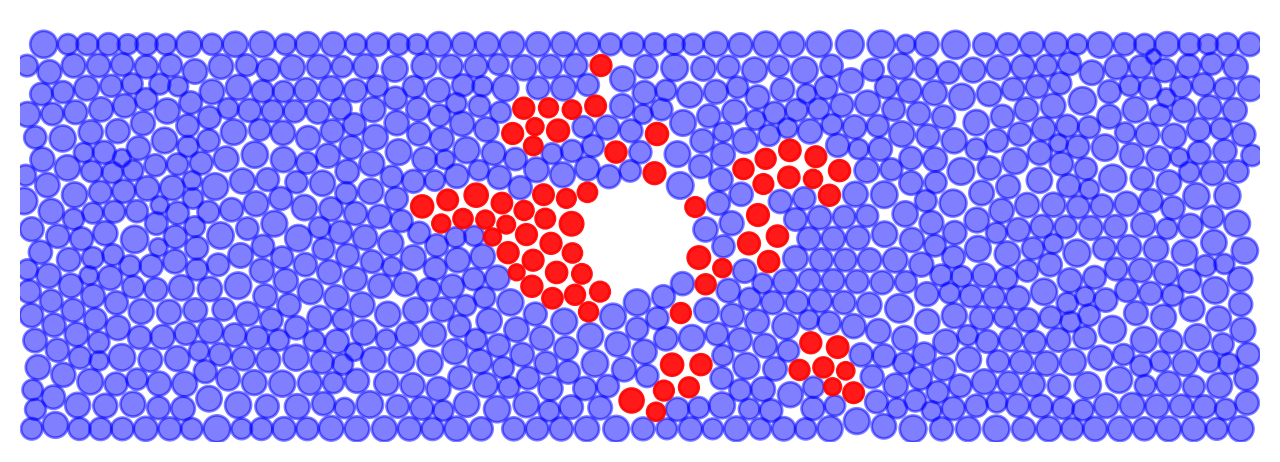}
				\end{subfigure}
				\\[4pt]
				
				\begin{subfigure}[t]{0.48\linewidth}
					\captionsetup{justification=raggedright,singlelinecheck=false,position=above}
					\caption{$\delta=0.001$: $\Delta {\bf r}^{\rm NA}$ for $\Delta t=900$}
					\hspace{-0.25\linewidth}\includegraphics[width=\linewidth]{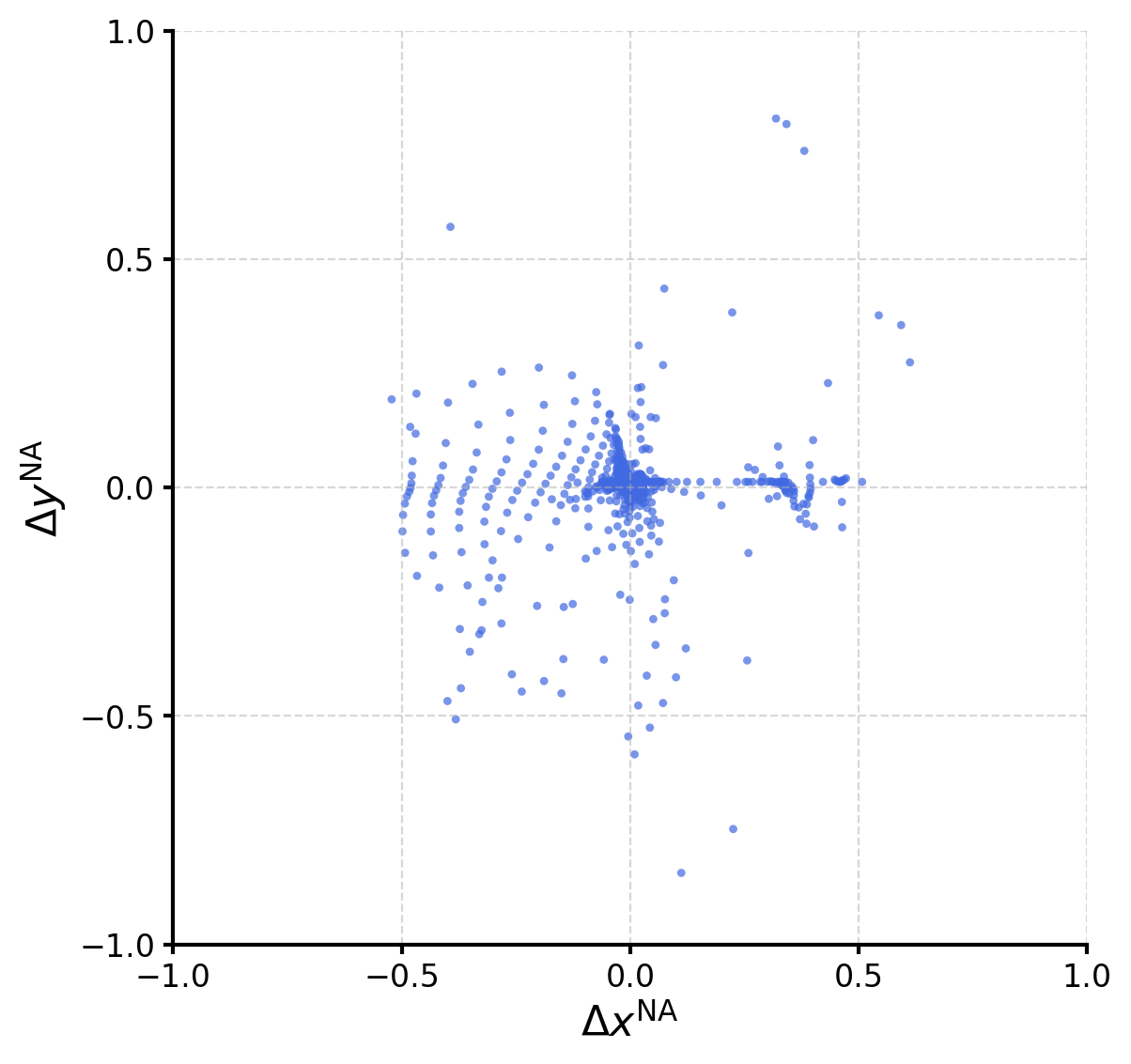}
				\end{subfigure}
				&
				\begin{subfigure}[t]{0.48\linewidth}
					\captionsetup{justification=raggedright,singlelinecheck=false,position=above}
					\caption{$\delta=0.15$: $\Delta {\bf r}^{\rm NA}$ for $\Delta t=900$}
					\hspace{-0.25\linewidth}\includegraphics[width=\linewidth]{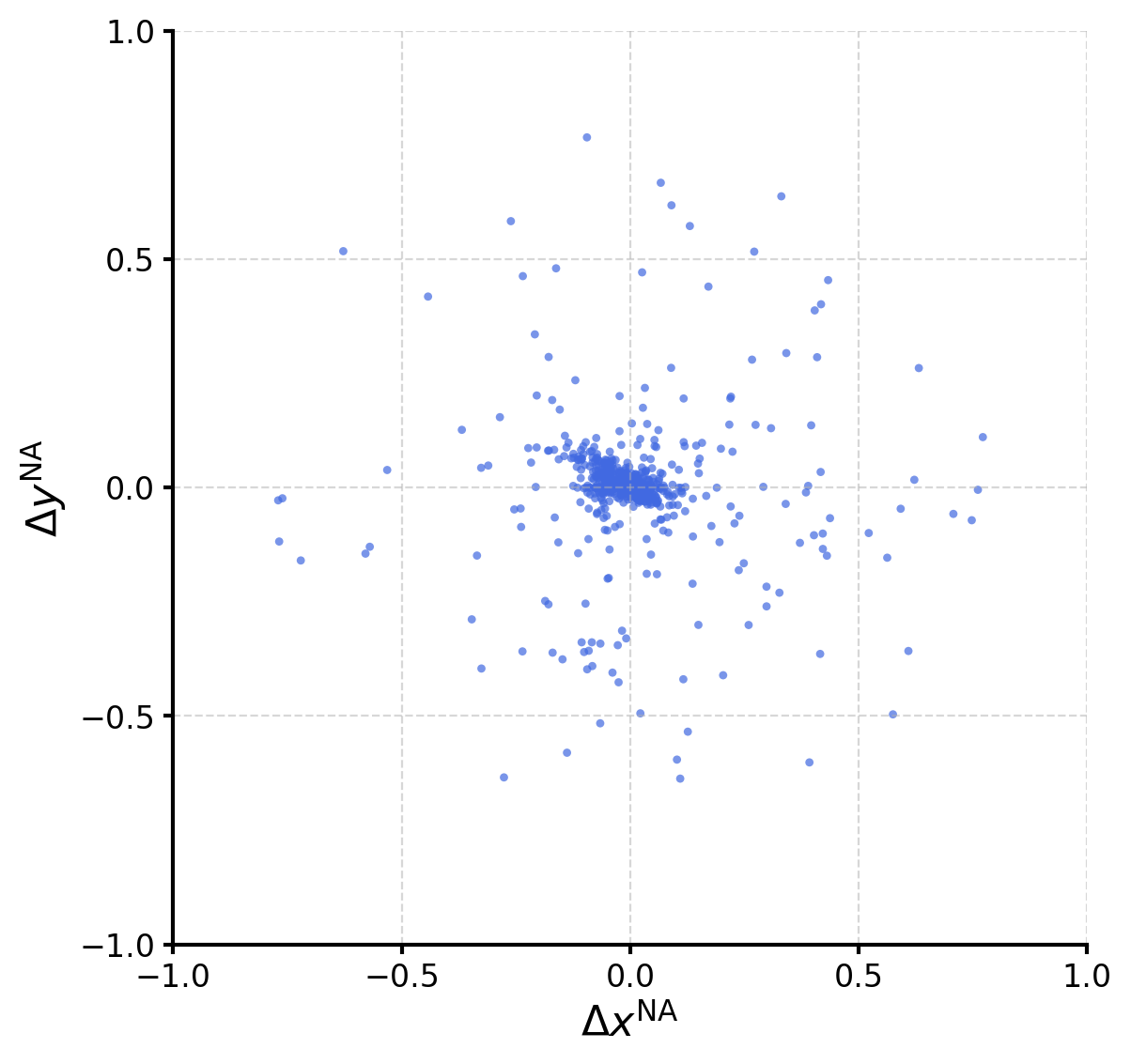}
				\end{subfigure}
			\end{tabular}
		}
		\captionsetup{justification=raggedright,singlelinecheck=false}
		\caption{Representative spatial flow profiles for $\delta = 0.001$ (a, c, e, g) and $\delta = 0.15$ (b, d, f, h). Shown are the displacement vector field $\Delta \mathbf{r}_i$ (a, b), the minimum non-affine squared displacement $D_{\rm min}^2$ (c, d), the neighbor change event indicator (e, f), and the non-affine displacement field $\Delta \mathbf{r}_i^{\rm NA}$ (g, h).
		}
		\label{fig:combined}
	\end{figure*}

	First, to characterize the spatial organization of plastic rearrangements, we compute the non-affine component of particle displacements between times $t$ and $t+\Delta t$. 
	For each particle $i$, the non-affine displacement is defined as
	\begin{equation}
		\Delta {\bf r}_i^{\rm NA} = \Delta {\bf r}_i - \overline{\bf v}\, \Delta t,
	\end{equation}
	where $\Delta {\bf r}_i = {\bf r}_i(t+\Delta t) - {\bf r}_i(t)$ and $\overline{\bf v}$ is the mean velocity of the system. 
	This subtraction removes the contribution from the global affine translation imposed by the external driving.
	In Fig.~\ref{fig:combined}, we show the non-affine displacement
	$\Delta \mathbf{r}_i^{\rm NA} = (\Delta x_i^{\rm NA}, \Delta y_i^{\rm NA})$
	for $\delta = 0.001$ (g) and $\delta = 0.15$ (h), together with the corresponding
	total displacements $\Delta \mathbf{r}_i$ (a, b), minimum non-affine squared
	displacement $D_{\rm min}^2$ (c, d), and neighbor-change-event indicator (e, f).
	The scatter plots of $(\Delta x_i^{\rm NA}, \Delta y_i^{\rm NA})$ reveal strongly
	directional and anisotropic rearrangements in the $\delta = 0.001$ case, in
	contrast to the nearly isotropic plastic activity observed for $\delta = 0.15$.
	
	We then quantify this anisotropy from the scatter plots.
	Using $\Delta {\bf r}_i^{\rm NA}=(\Delta x_i^{\rm NA},\Delta y_i^{\rm NA})$, we compute a tensor $M$, given by
	\begin{eqnarray}
		M =
		\begin{pmatrix}
			\dfrac{1}{N}\sum\limits_{i=1}^{N} (\Delta x_i^{\rm NA})^2 
			&
			\dfrac{1}{N}\sum\limits_{i=1}^{N} \Delta x_i^{\rm NA}\Delta y_i^{\rm NA}
			\\[8pt]
			\dfrac{1}{N}\sum\limits_{i=1}^{N} \Delta x_i^{\rm NA}\Delta y_i^{\rm NA}
			&
			\dfrac{1}{N}\sum\limits_{i=1}^{N} (\Delta y_i^{\rm NA})^2
		\end{pmatrix} . \nonumber \\
	\end{eqnarray}
	The tensor $M$ has two eigenvalues, which we denote by $\lambda_{\rm max}$ (larger) and $\lambda_{\rm min}$ (smaller). 
	We define an anisotropy parameter $A$ as
	\begin{equation}
		A = \frac{\lambda_{\rm max} - \lambda_{\rm min}}{\lambda_{\rm max} + \lambda_{\rm min}}.
	\end{equation}
	A value $A \approx 1$ indicates strongly directional, sliding-like rearrangements, while $A \approx 0$ corresponds to isotropic rearrangements.
	
\begin{figure}[htbp]
		\centering
		\includegraphics[width=0.98\linewidth]{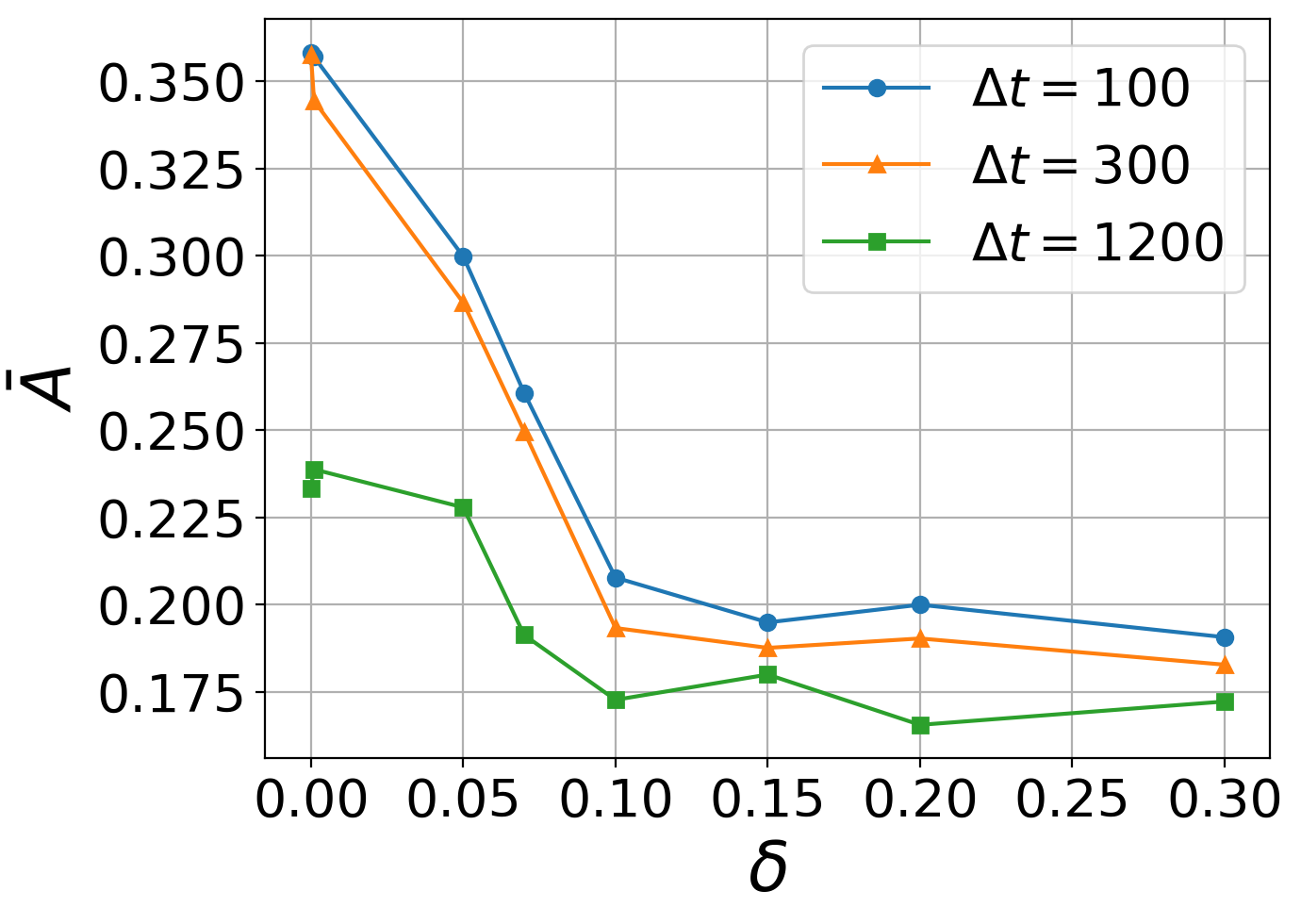}
		\captionsetup{justification=raggedright,singlelinecheck=false}
		\caption{Averaged anisotropy parameter $\overline A$ as a function of polydispersity $\delta$
			for different time intervals $\Delta t$.
		}
		\label{fig:Avsdelta}
	\end{figure}

	In Fig.~\ref{fig:Avsdelta}, we show the anisotropy parameter $A$ averaged over
	many snapshots, denoted by $\overline{A}$, as a function of the polydispersity
	$\delta$ for different time intervals $\Delta t$.
	At short time intervals, we observe a large anisotropy,
	$\overline{A} \approx 0.35$ for $\delta \approx 0$, which decreases
	systematically with increasing $\delta$ and reaches an approximately constant
	plateau for $\delta \gtrsim 0.1$.
	This behavior is consistent with the visual observations discussed above, but it is here confirmed in a quantitative and statistical manner. Interestingly, the results indicate convergence toward the heterogeneous plasticity typical of disordered materials for a polydispersity value of $\delta \approx 0.1$. This value therefore plays the role of a critical, or threshold, polydispersity below which crystalline-like behavior becomes dominant.

	As the time interval $\Delta t$ is increased, the value of $\overline{A}$ near
	$\delta \approx 0$ is progressively reduced.
	This reduction reflects the fact that multiple plastic rearrangements occurring
	over long times smear out the initially directional motion, yielding a more
	isotropic overall pattern.

	\section{Results: Effect of External Driving}
	\label{sec:external_driving}
	
\begin{figure*}[htbp]
		\centering
		\begin{subfigure}{0.32\textwidth}
			\captionsetup{justification=raggedright, singlelinecheck=false, position=above}
			\caption{$f^{\rm ext}=10^{-3}$, $\Delta t=300$.}
			\includegraphics[width=\textwidth]{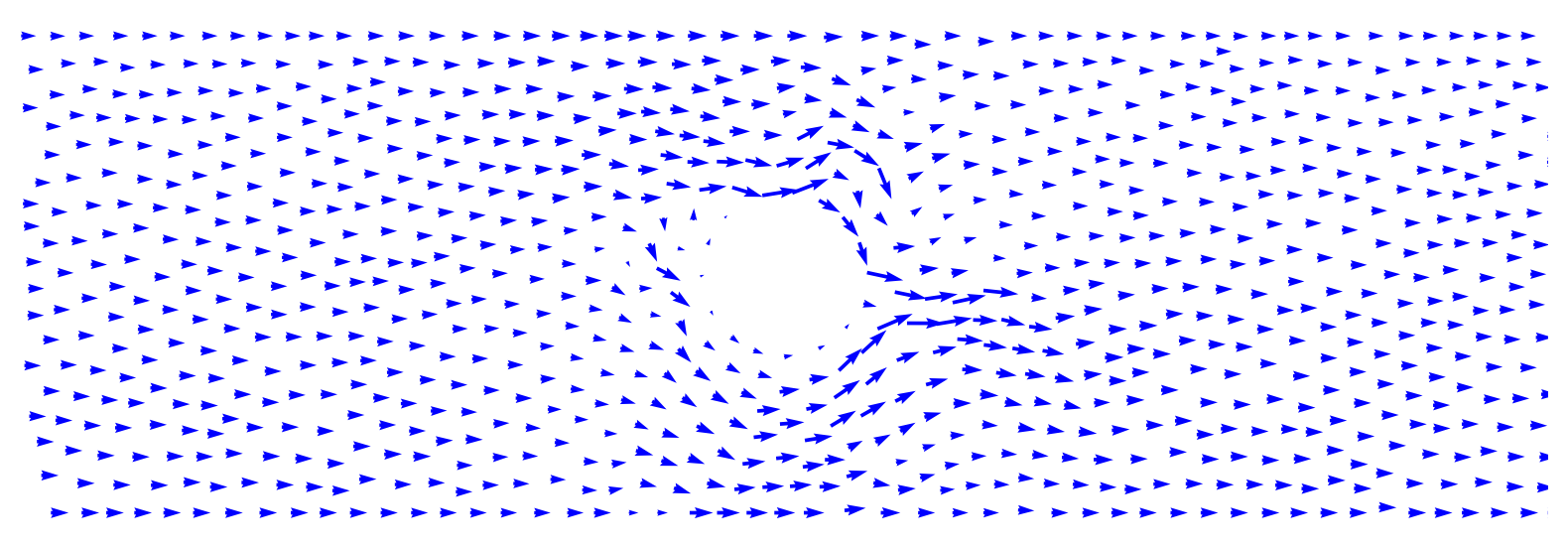}
			\label{fig:a}
		\end{subfigure}
		\hfill
		\begin{subfigure}{0.32\textwidth}
			\captionsetup{justification=raggedright, singlelinecheck=false, position=above}
			\caption{$f^{\rm ext}=10^{-2}$, $\Delta t=300$.}
			\includegraphics[width=\textwidth]{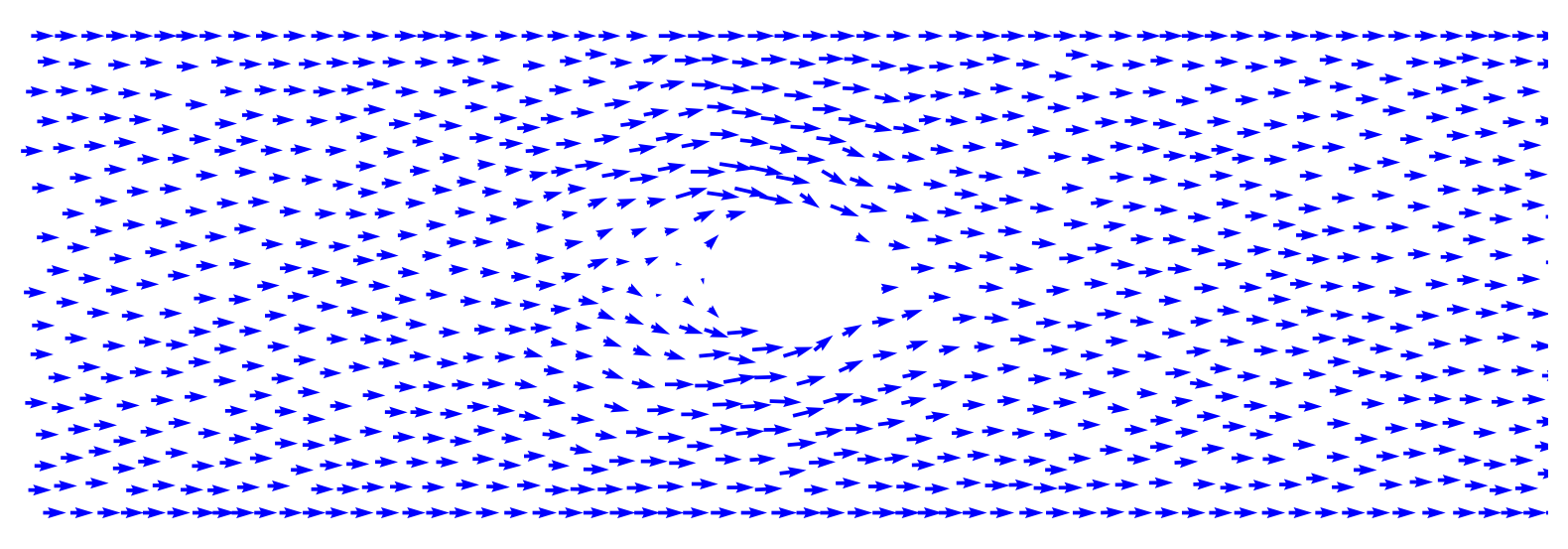}
			\label{fig:b}
		\end{subfigure}
		\hfill
		\begin{subfigure}{0.32\textwidth}
			\captionsetup{justification=raggedright, singlelinecheck=false, position=above}
			\caption{$f^{\rm ext}=10^{-1}$, $\Delta t=50$.}
			\includegraphics[width=\textwidth]{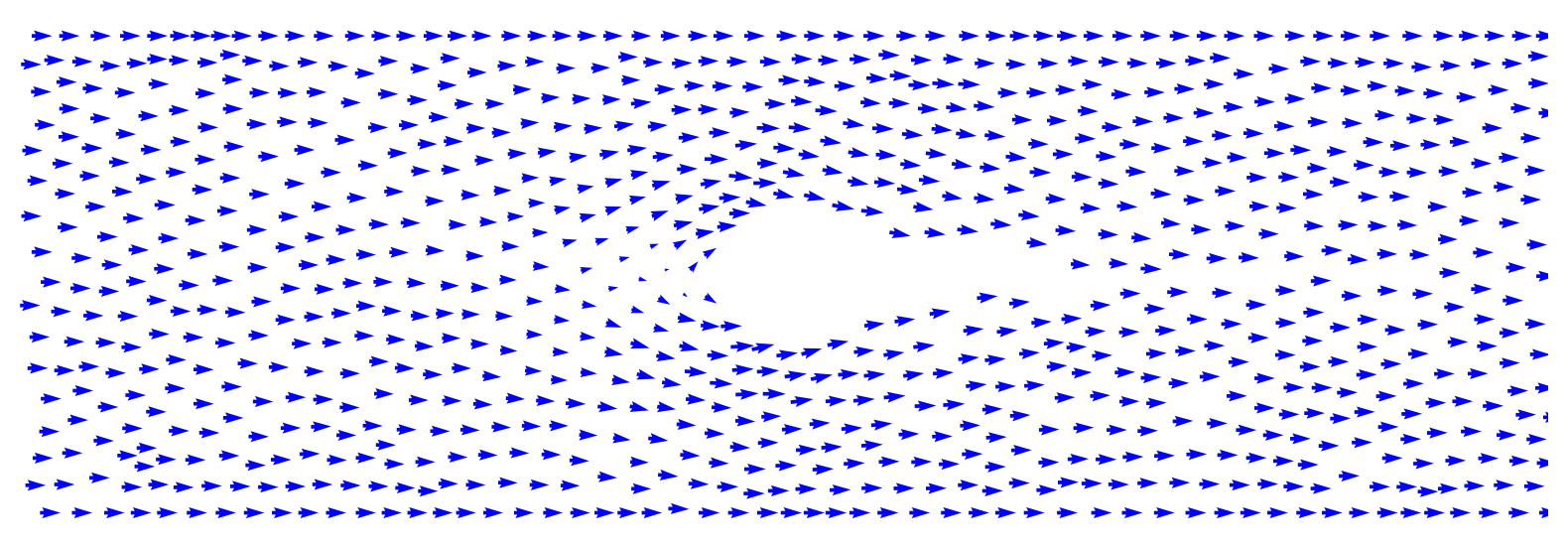}
			\label{fig:c}
		\end{subfigure}
		\captionsetup{justification=raggedright,singlelinecheck=false}
		\caption{(a-c): Representative plots of the displacement vectors $\Delta \mathbf{r}_i$ for different values of $f^{\rm ext}$. The time interval used to compute the displacements is $\Delta t = 300$ in (a), $300$ in (b), and $50$ in (c), respectively. For visual clarity, the displacement vectors are rescaled by factors of $2$ in (a), $0.25$ in (b), and $0.125$ in (c).}
		\label{fig:force}
	\end{figure*}

	We now investigate the effect of external driving, focusing on the amorphous
	system with $\delta = 0.15$.
	In Fig.~\ref{fig:force}, we show the displacement vector field
	$\Delta \mathbf{r}_i$ for different magnitudes of the driving force, ranging from
	weak to strong forcing.
	As discussed in the previous section, the displacement field is highly
	heterogeneous at weak driving.
	With increasing $f^{\rm ext}$, the displacement vectors progressively align
	along the $x$ direction, as expected for a flow-dominated regime.
	At very large $f^{\rm ext}$, an empty region appears on the downstream side of
	the obstacle, indicating that the driving time scale becomes much shorter than
	the relaxation time associated with particle rearrangements mediated by
	interparticle interactions.

	We further investigate the effect of driving magnitude and find an
	interesting phenomenon in the very weak driving regime.
	For example, for $f^{\rm ext} = 10^{-4}$ at packing fraction $\phi = 1.2$, the
	dynamics completely stops after an initial transient period, despite the
	continuous application of the external force.
	This indicates that the central obstacle acts as a pinning center, i.e., particles are
	mechanically coupled through short-range interactions, forming force chains that
	span the system and prevent flow.
	In other words, a sufficiently large external force is required to induce sustained flow. This is the phenomenon of yield drag, which has been observed experimentally in both two-dimensional~\cite{Raufaste2007} and three-dimensional~\cite{Cantat2006} foams, as well as in numerical simulations of two-dimensional foams based on the Surface Evolver and Potts models~\cite{Raufaste2007}. 
	To the best of our knowledge, we report here the first observation of yield-drag phenomena in molecular dynamics simulations.
	This behavior is also reminiscent of the yielding transition in amorphous materials, where flow occurs only when the applied stress exceeds a critical threshold (the yield stress), while the system remains arrested below this threshold~\cite{bonn2017yield}, as well as of depinning transitions in driven disordered systems, in which sustained motion sets in only when the applied force exceeds a critical value, whereas the system remains pinned below this threshold~\cite{fisher1998collective,reichhardt2016depinning}.

	To study this phenomenon quantitatively, we perform many independent simulations with different random initial configurations and determine whether each run reaches a steady flowing state or becomes completely arrested. A simulation is classified as arrested if the total potential energy becomes strictly constant at long times ($t = 10^5$). For each combination of parameters, we perform 100 independent realizations for different values of $f^{\rm ext}$ at packing fractions $\phi = 0.9$, $1.0$, $1.05$, $1.1$, $1.15$, $1.2$, and $1.3$.

\begin{figure}[htbp]
		\includegraphics[width=0.9\linewidth]{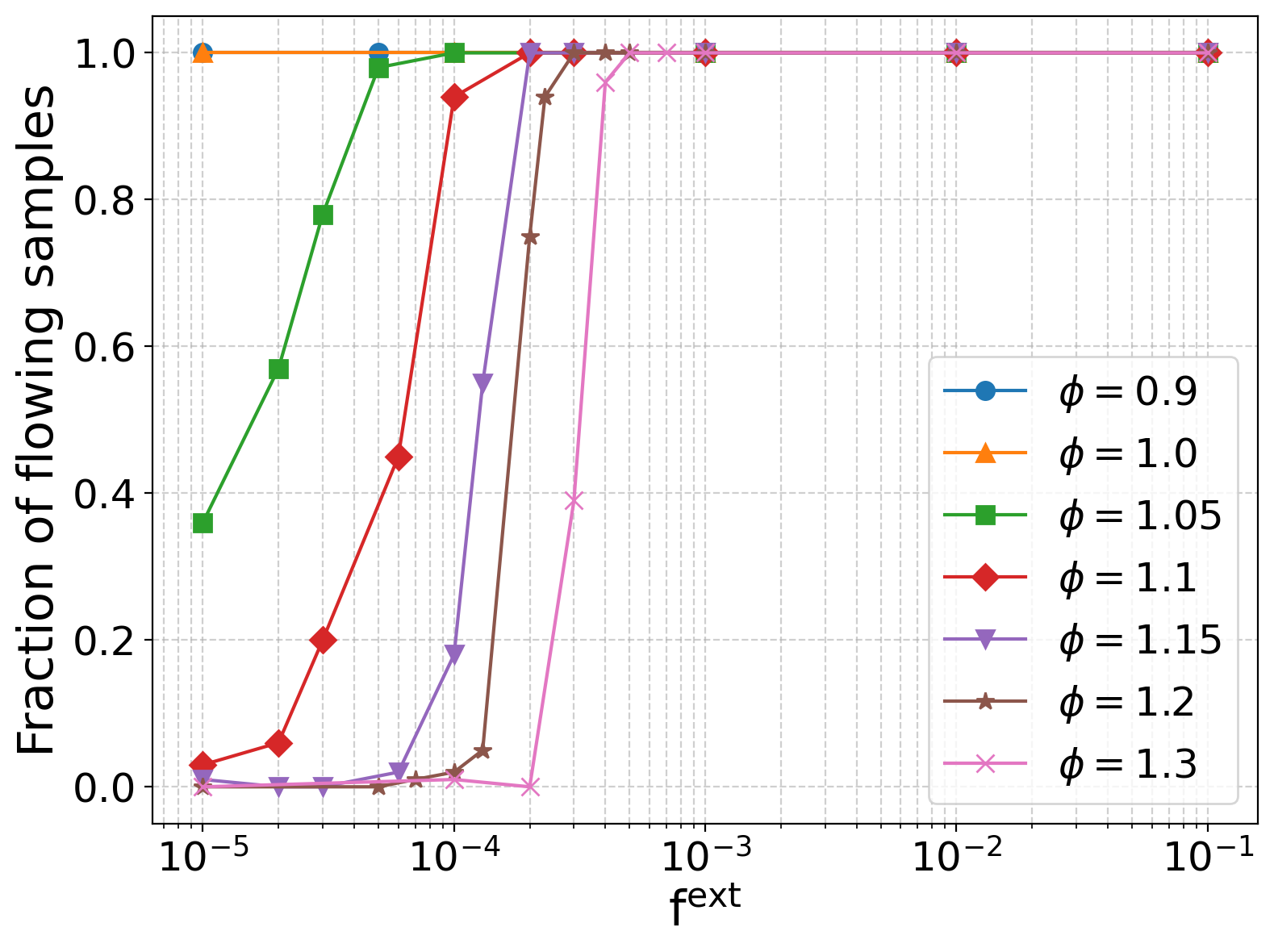}
		\captionsetup{justification=raggedright,singlelinecheck=false}
		\caption{Fraction of flowing samples among 100 independent runs as a function of $f^{\rm ext}$ for several packing fractions $\phi$.
		}
		\label{fig:flow_probability}
	\end{figure}

	We then measure the fraction of runs that reach steady-state flow, as shown in Fig.~\ref{fig:flow_probability}.
	For lower packing fractions, $\phi = 0.9$ and $1.0$, all simulations reach steady flow within our simulation window down to $f^{\rm ext} = 10^{-5}$.
	In contrast, for $\phi = 1.05, 1.1$, a finite fraction of samples becomes arrested at lower values of $f^{\rm ext}$.
	Upon further increasing the packing fraction to $\phi = 1.15$ and above, we observe a sharper, step-like transition, which allows us to identify a critical driving force (or yield drag) $f_c^{\rm ext} \approx 5\times 10^{-5}$ for $\phi = 1.15$, $f_c^{\rm ext} \approx 10^{-4}$ for $\phi = 1.2$, and $f_c^{\rm ext} \approx 2 \times 10^{-4}$ for $\phi = 1.3$. Hence, in this system, a nonzero yield drag $f_c^{\rm ext}$ emerges around $\phi \approx 1.15$, and $f_c^{\rm ext}$ increases with increasing $\phi$.
	
	It is worth noting that this threshold value $\phi \approx 1.15$ is higher than the close-packing fraction $\phi \approx 0.91$ and the (disordered) jamming transition point $\phi_J \approx 0.85$ for real two-dimensional foams. This discrepancy is a known feature of the bubble model \cite{durian1995foam}:
	the packing fraction of the bubble model corresponds to effectively lower values in real systems, due to the possibility of strong particle overlaps inherent to the model.
	Our results therefore suggest a possible route to establish a mapping between the packing fraction of the bubble model and that of real foams, based on the measured values of yield drag.

	The progressive sharpening of the flow fraction seen in Fig.~\ref{fig:flow_probability} and the emergence of a nonzero $f_c^{\rm ext}$ with increasing packing fraction are reminiscent of the yielding transition observed in jammed materials under rheological measurements (see, e.g., Fig.~3 of Ref.~\cite{bonn2017yield}). The crossover regime, where the flow fraction takes intermediate values between 0 and 1 for $\phi = 1.05$–$1.1$, may be attributed to several effects. First, intermittency may play a role: in some realizations, the system can form force-chain configurations that effectively resist the external drive. Second, finite-size effects may be important, as smaller systems exhibit stronger sample-to-sample fluctuations. Finally, limitations due to the finite simulation time may also contribute, since the classification of a state as flowing or arrested is necessarily based on observations over a finite time window. A more systematic investigation of this phenomenon, including critical properties, finite-size effects, and dependence on polydispersity, is left for future work.

	\section{Discussion: relevance for foams}
	\label{sec:foam}
	
	Since our numerical study is inspired by experiments on foams~\cite{dollet2007two}, it is important to assess the extent to which our simulation model, which is closely related to Durian's bubble model~\cite{durian1995foam,Durian1997}, provides a realistic description of foam flows and to identify the essential physical ingredients that it may neglect. Indeed, this model seems at first sight less realistic than vertex models \cite{Okuzono1995,cantat2011gibbs}, or Surface Evolver simulations \cite{Boulogne2011}, which account explicitly for films and/or vertices between bubbles. Moreover, Surface Evolver simulations have been extended into the so-called ``viscous froth model'' \cite{Kern2004} to account for the external friction force arising from the sliding of films against confining walls, which makes it much more realistic.
	
	However, despite its apparent oversimplicity, our simulation model retains the same essential ingredients to capture flows of confined foams, with a big advantage in terms of computational cost, which is precisely why it was devised in the first place \cite{durian1995foam}. First, even if surface tension is not directly present in the simulation model, its mechanical effect emerges when the particles are sufficiently confined. In particular, the bubble model is known to exhibit a nonzero shear modulus, and a nonzero yield stress, above a critical packing fraction \cite{durian1995foam,Durian1997}. This is because the repulsive potential in Eq.~(\ref{Eq:potential}) between overlapping particles mimics the repulsive potential associated with bubbles deformed with flat facets, even though the functional form of the potential might differ \cite{Morse1993,Hohler2019}. More precisely, the energy scale $\epsilon$ appearing in Eq.~(\ref{Eq:potential}) can be mapped, up to a multiplicative prefactor, to the product of surface tension and of the square of a characteristic length scale of the bubble (e.g. their mean radius).  Second, the term $\zeta d\mathbf{r}_i(t)/dt$ in the overdamped dynamics of Eq.~(\ref{eq:overdamped_dynamics}) can be seen as a friction force of the particles sliding fictitious fixed walls confining the assembly of particles in their plane. As such, this term is in close analogy to the force added to the viscous forth model, except that it is applied to the particle centroids and not to films. Hence, it is reasonable to expect, and indeed it is seen in our study, that most of the confined foam flow phenomenology (yield stress, plastic rearrangements, dissipative effects) is reproduced by the bubble model.
	
	Probably the main missing ingredient is the absence of bubble/bubble friction as they slide past each other. Indeed, it has been shown to be the key physical ingredient to explain the viscous rheological properties of unconfined 3D foams \cite{Denkov2008}. It is therefore a natural perspective of our work to extend the model by adding a friction term dependent on the relative velocity of neighboring particles.  However, the current simulation model is probably still fine for most 2D confined flows, where ``external'' foam-wall friction often dominates ``internal'' bubble-bubble friction \cite{Raufaste2009,BenSalem2013}.

	\section{Conclusion and Perspectives}
	\label{sec:conclusion}

	Conventionally, the mechanical response of materials has been studied primarily under simple deformation protocols such as shear, uniaxial tension, or compression, in order to investigate the mechanisms of plasticity as well as practical aspects relevant to rheology and mechanical engineering.
	
	Confined channel flow around an obstacle provides an alternative and complementary geometry to probe plasticity, rheology, and yielding behavior. This configuration introduces additional ingredients, such as geometric confinement and interactions with an obstacle, which are absent in standard bulk deformation protocols. These effects are not only of fundamental interest, recalling classical problems such as Stokes flow in hydrodynamics, but are also of significant practical relevance.
	
	Taking advantage of the flexibility of molecular simulations, we perform extensive calculations to explore a wide range of parameter space, in particular the polydispersity, the magnitude of the external driving force, and the packing fraction. We identify a threshold value of the polydispersity index that marks a crossover between two distinct regimes: highly directional, sliding-like plastic motion characteristic of crystalline systems, and more isotropic rearrangements typical of amorphous materials. In addition, we observe the existence of a critical external force above which the system reaches a steady flowing state and below which it remains arrested, a phenomenon known as yield drag.

	Interestingly, this yield-drag behavior emerges only above a threshold packing fraction, reminiscent of the rheology of yield-stress fluids, where the onset of flow is controlled by packing density. The appearance of yield drag as a function of packing fraction suggests a possible route to map the packing fraction of the bubble model in simulations onto that of real foams in laboratory experiments composed of non-overlapping bubbles.

	Our study opens several directions for future investigations. First, it would be valuable to explore more systematically the effects of geometric confinement and obstacle size, which were kept fixed in the present work. It would also be important to examine more carefully the nature of the yield-drag phenomenon, in particular to determine whether it corresponds to a genuine transition with well-defined critical properties~\cite{lin2014scaling,nicolas2018deformation}, or merely to a smooth crossover. Addressing this question will require systematic finite-size scaling analyses based on extensive simulations with varying system size, geometry, and the number of particles.
	
	Moreover, the prediction of future heterogeneous dynamics or plastic activity in amorphous systems from static structural snapshots using machine learning techniques has recently become an active area of research~\cite{richard2020predicting,jung2025roadmap}. Such approaches have so far been applied mostly to bulk, spatially uniform systems. The confined channel flow around an obstacle studied in this paper provides a new and challenging test case for machine-learning-based prediction, as the presence of confinement and obstacles will require additional descriptors of the local structural environment.

	\begin{acknowledgments}
		We thank the support by MIAI@Grenoble Alpes and the Agence Nationale de la Recherche under France 2030 with the reference ANR-23-IACL-0006. 
		This work was also supported by LabEx TEC21/UGA through the French National Research Agency in the framework of the ``France 2030'' program (ANR-15-IDEX-02).
	\end{acknowledgments}
	
	\section*{Data Availability}
	
	All the source codes and dataset used in this paper are openly available at
	\href{https://github.com/mazloum-bahaa}
	{https://github.com/mazloum-bahaa}.

	\appendix

	\section{Minimum non-affine squared displacement $D^2_{\rm min}$}
	\label{sec:D2min}
	
	We compute the minimum non-affine squared displacement $D^2_{\rm min}$~\cite{falk1998dynamics}.
	$D^2_{\rm min}$ is defined as
	\begin{eqnarray}
		D^2_{{\rm min}, i} &=& \frac{1}{n_i} \sum_{j \in \mathcal{N}_i} | ({\bf r}_j(t+\Delta t)-{\bf r}_i(t+\Delta t)) \\ \nonumber 
		&\qquad& \qquad \qquad  - (I+E^*)({\bf r}_j(t)-{\bf r}_i(t))|^2 ,
	\end{eqnarray}
	where $n_i$ is the number of neighbor particles of particle $i$, 
	$\mathcal{N}_i$ denotes the set of neighbors of particle $i$, 
	$I$ is the identity matrix, and 
	$E^*$ is the matrix representing the best-fit local affine deformation. 
	Thus, $D^2_{{\rm min},i}$ captures only the non-affine contribution to particle motion, 
	while the affine deformation associated with $I+E^*$ is subtracted.
	We define the neighbors of particle $i$ as the particles located within a cutoff distance $r_{\rm min}$ at time $t$, where $r_{\rm min}$ is chosen as the position of the first minimum of the radial distribution function.

	\section{Detecting neighbor change events}
	\label{sec:T1}

	In this paper, we define a neighbor change indicator for each particle as a binary variable that equals $1$ if a neighbor change rearrangement occurs and $0$ otherwise.
	
	For each particle, we compute its list of neighbors at times $t$ and $t+\Delta t$ and compare the two lists. If the neighbor lists are identical, no neighbor change event is detected. If the lists differ, the particle is classified as undergoing a neighbor change rearrangement. At the microscopic level, plastic activity modifies the neighbor list through two distinct mechanisms: (i) bond breaking and (ii) bond formation. In our study, these two mechanisms are measured independently and do not necessarily occur in correlated pairs, in contrast to the T1 events encountered in dry two-dimensional foams. Indeed, a genuine T1 event consists of the simultaneous swapping of neighbors among four adjacent bubbles, so that bond breaking and bond formation necessarily occur together in matched pairs.
	
	{\it (i) Bond breaking}:
	A bond-breaking event occurs when one or more particles that belong to the neighbor list at time $t$ disappear from the list at time $t+\Delta t$.
	To robustly identify this process and reduce noise due to local fluctuations, we define the neighbor lists using two slightly different cutoff distances~\cite{nishikawa2022relaxation,takaha2025avalanche}: $r_1 = r_{\rm min} - \delta r/2$ at time $t$ and $r_2 = r_{\rm min} + \delta r/2$ at time $t+\Delta t$,
	where $r_{\rm min}$ is the position of the first minimum of the radial distribution function.
	We set $\delta r=0.2$.
	The introduction of a small margin $\delta r$ suppresses spurious changes in the neighbor list and allows us to detect genuine bond-breaking events.
	
	{\it (ii) Bond formation}:
	A bond-formation event occurs when one or more particles that were not neighbors at time $t$ enter the neighbor list at time $t+\Delta t$. To detect this process, we reverse the asymmetric cutoffs and define $r_1 = r_{\rm min} + \delta r/2$ at time $t$ and $r_2 = r_{\rm min} - \delta r/2$ at time $t+\Delta t$.
	
	Finally, the neighbor change indicator for particle $i$ is defined to be $1$ if either bond breaking or bond formation is detected between times $t$ and $t+\Delta t$, and $0$ otherwise.

	\section{Sliding rearrangements}
	\label{sec:sliding}
	
	We present additional samples showing directional, sliding-like motion in the crystalline system with $\delta = 0.001$, as characterized by the binary neighbor change indicator, in addition to those shown in Fig.~\ref{fig:delta=0.001}. We also present one sample showing a plastic event occurring away from the obstacle.

	\begin{figure}[htbp]
		\begin{subfigure}[t]{0.98\linewidth}
			\captionsetup{justification=raggedright, singlelinecheck=false, position=above}
			\includegraphics[width=\linewidth]{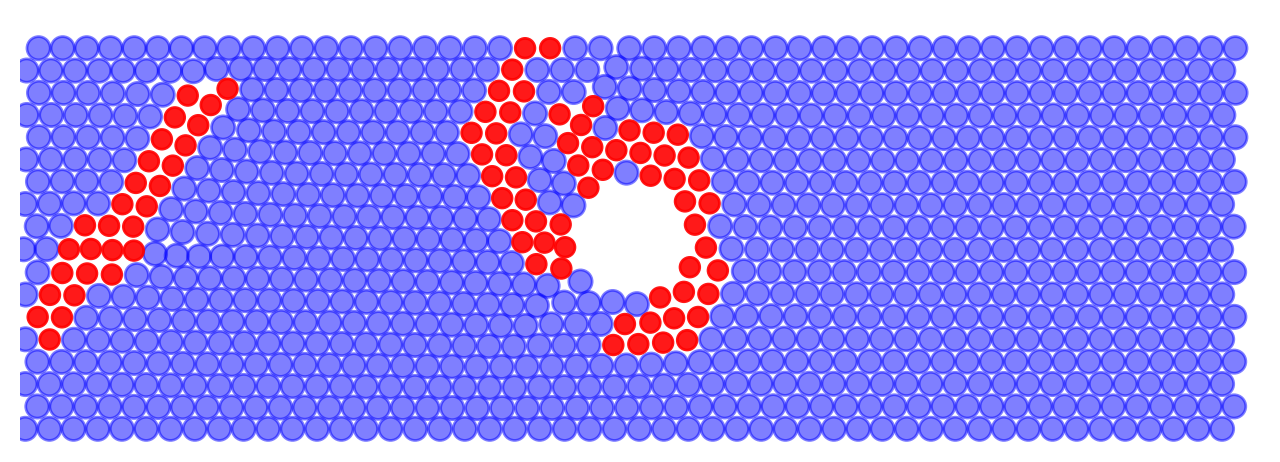}
			\label{fig:xxx}
		\end{subfigure}
		\hfill
		\begin{subfigure}[t]{0.98\linewidth}
			\captionsetup{justification=raggedright, singlelinecheck=false, position=above}
			\includegraphics[width=\linewidth]{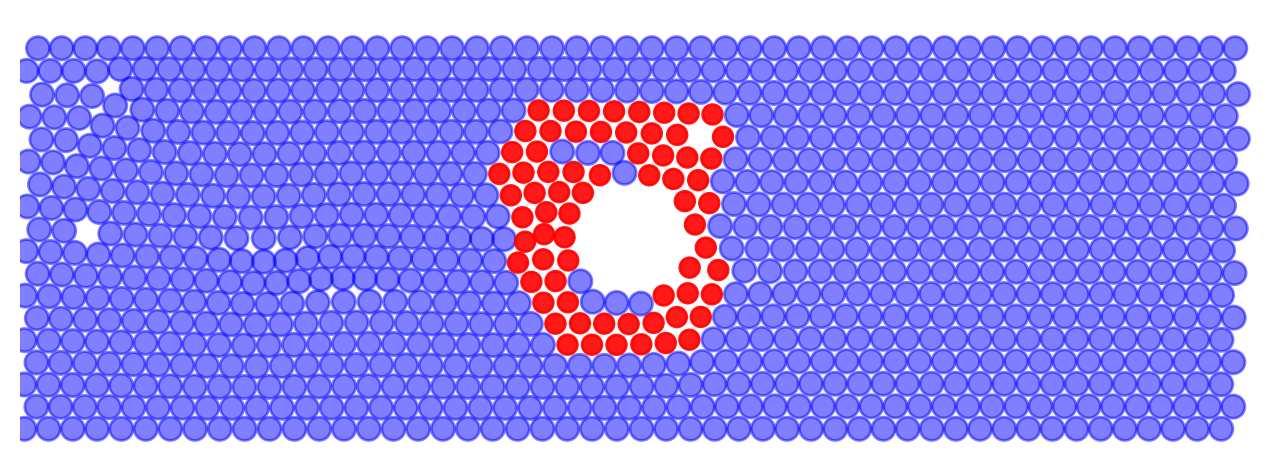}
			\label{fig:xxx}
		\end{subfigure}
		\hfill
		\begin{subfigure}[t]{0.98\linewidth}
			\captionsetup{justification=raggedright, singlelinecheck=false, position=above}
			\includegraphics[width=\linewidth]{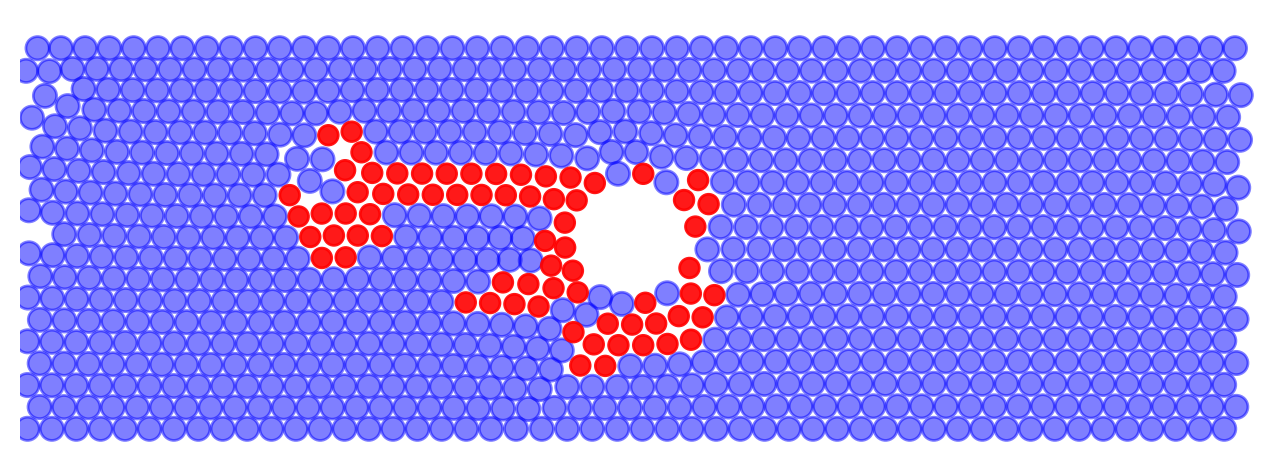}
			\label{fig:xxx}
		\end{subfigure}
		\captionsetup{justification=raggedright,singlelinecheck=false}
		\caption{Three representative samples showing sliding-like motion in the crystalline-like system with $\delta = 0.001$, characterized by neighbor change events for $\Delta t = 1200$. }
		\label{fig:sliding}
	\end{figure}

	\bibliography{refs_foam_projects}

@article{dollet2007two,
  title={Two-dimensional flow of foam around a circular obstacle: local measurements of elasticity, plasticity and flow},
  author={Dollet, Benjamin and Graner, Fran{\c{c}}ois},
  journal={J. Fluid Mech.},
  volume={585},
  pages={181--211},
  year={2007},
  publisher={Cambridge University Press}
}

@article{durian1995foam,
  title={Foam mechanics at the bubble scale},
  author={Durian, Douglas J},
  journal={Phys. Rev. Lett.},
  volume={75},
  number={26},
  pages={4780},
  year={1995},
  publisher={APS}
}

@article{bonn2017yield,
  title={Yield stress materials in soft condensed matter},
  author={Bonn, Daniel and Denn, Morton M and Berthier, Ludovic and Divoux, Thibaut and Manneville, S{\'e}bastien},
  journal={Rev. Mod. Phys.},
  volume={89},
  number={3},
  pages={035005},
  year={2017},
  publisher={APS}
}

@article{nicolas2018deformation,
  title={Deformation and flow of amorphous solids: Insights from elastoplastic models},
  author={Nicolas, Alexandre and Ferrero, Ezequiel E and Martens, Kirsten and Barrat, Jean-Louis},
  journal={Rev. Mod. Phys.},
  volume={90},
  number={4},
  pages={045006},
  year={2018},
  publisher={APS}
}

@article{maloney2006amorphous,
  title={Amorphous systems in athermal, quasistatic shear},
  author={Maloney, Craig E and Lemaitre, Anael},
  journal={Phys. Rev. E},
  volume={74},
  number={1},
  pages={016118},
  year={2006},
  publisher={APS}
}

@article{falk1998dynamics,
  title={Dynamics of viscoplastic deformation in amorphous solids},
  author={Falk, Michael L and Langer, James S},
  journal={Phys. Rev. E},
  volume={57},
  number={6},
  pages={7192},
  year={1998},
  publisher={APS}
}

@article{nishikawa2022relaxation,
  title={Relaxation dynamics in the energy landscape of glass-forming liquids},
  author={Nishikawa, Yoshihiko and Ozawa, Misaki and Ikeda, Atsushi and Chaudhuri, Pinaki and Berthier, Ludovic},
  journal={Phys. Rev. X},
  volume={12},
  number={2},
  pages={021001},
  year={2022},
  publisher={APS}
}

@article{takaha2025avalanche,
  title={Avalanche criticality emerges by thermal fluctuation in a quiescent glass},
  author={Takaha, Yuki and Mizuno, Hideyuki and Ikeda, Atsushi},
  journal={Phys. Rev. E},
  volume={112},
  number={4},
  pages={L043401},
  year={2025},
  publisher={APS}
}

@book{Stevenson2012,
    author = {P. Stevenson},
    title = {Foam Engineering: Fundamentals and Applications},
	publisher = {John Wiley \& Sons, Ltd},
    editor = {Wiley},
    year = {2012}
}

@article{Ma2012,
	title = {Visualization of improved sweep with foam in heterogeneous porous media using microfluidics},
	volume = {8},
	journal = {Soft Matter},
	author = {K. Ma and R. Liontas and C. A. Conn and G. J. Hirasaki and S. L. Biswal},
	year = {2012},
	pages = {10669--10675}
}

@article{Geraud2016,
	title = {The flow of a foam in a two-dimensional porous medium},
	volume = {52},
	journal = {Water Resour. Res.},
	author = {B. Géraud and S. A. Jones and I. Cantat and B. Dollet and Y. Méheust},
	year = {2016},
	pages = {773--790}
}

@article{Katgert2008,
	title = {Rate dependence and role of disorder in linearly sheared two-dimensional foams},
	volume = {101},
	journal = {Phys. Rev. Lett.},
	author = {G. Katgert and M. E. Möbius and M. van Hecke},
	year = {2008},
	pages = {058301}
}

@book{Oswald2014,
    author = {P. Oswald},
    title = {Rheophysics: The Deformation and Flow of Matter},
	publisher = {Cambridge University Press},
    year = {2014}
}

@article{weaire1984soap,
  title={Soap, cells and statistics—random patterns in two dimensions},
  author={Weaire, Denis and Rivier, Nicolas},
  journal={Contemporary Physics},
  volume={25},
  number={1},
  pages={59--99},
  year={1984},
  publisher={Taylor \& Francis}
}

@article{Boulogne2011,
	title = {Elastoplastic flow of foam around an obstacle},
	volume = {83},
	journal = {Phys. Rev. E},
	author = {F. Boulogne and S. J. Cox},
	year = {2011},
	pages = {041404}
}

@article{Raufaste2007,
	title = {Yield drag in a two-dimensional foam flow around a circular obstacle: Effect of liquid fraction},
	volume = {23},
	journal = {Eur. Phys. J. E},
	author = {C. Raufaste and B. Dollet and S. J. Cox and Y. Jiang and F. Graner},
	year = {2007},
	pages = {217--228}
}

@article{Viitanen2019,
	title = {Probing the local response of a two-dimensional foam},
	volume = {92},
	journal = {Eur. Phys. J. B},
	author = {L. Viitanen and J. Koivisto and A. Puisto and M. Alava and S. Santucci},
	year = {2019},
	pages = {38}
}

@article{Bragg1947,
	title = {A dynamical model of a crystal structure},
	volume = {190},
	journal = {Proc. Roy. Soc. A},
	author = {L. Bragg and J. F. Nye},
	year = {1947},
	pages = {474--481}
}

@article{Argon1979,
	title = {Plastic flow in a disordered bubble raft (an analog of a metallic glass)},
	volume = {39},
	journal = {Mater. Sci. Eng.},
	author = {A. S. Argon and H. Y. Kuo},
	year = {1979},
	pages = {101--109}
}

@article{Bulatov1994,
	title = {A stochastic model for continuum elasto-plastic behavior. I. Numerical approach and strain localization},
	volume = {2},
	journal = {Model. Simul. Mater. Sci. Eng.},
	author = {V. V. Bulatov and A. S. Argon},
	year = {1994},
	pages = {167--184}
}

@article{ghimenti2024shear,
  title={Shear-induced phase behavior and topological defects in two-dimensional crystals},
  author={Ghimenti, Federico and Ozawa, Misaki and Biroli, Giulio and Tarjus, Gilles},
  journal={Phys. Rev. B},
  volume={109},
  number={10},
  pages={104114},
  year={2024},
  publisher={APS}
}

@article{richard2020predicting,
  title={Predicting plasticity in disordered solids from structural indicators},
  author={Richard, David and Ozawa, Misaki and Patinet, Sylvain and Stanifer, Ethan and Shang, B and Ridout, Sean A and Xu, Bin and Zhang, Ge and Morse, PK and Barrat, J-L and others},
  journal={Physical Review Materials},
  volume={4},
  number={11},
  pages={113609},
  year={2020},
  publisher={APS}
}

@article{Cantat2006,
	title = {Stokes experiment in a liquid foam},
	volume = {18},
	journal = {Phys. Fluids},
	author = {I. Cantat and O. Pitois},
	year = {2006},
	pages = {083302}
}

@article{sethna2017deformation,
  title={Deformation of crystals: Connections with statistical physics},
  author={Sethna, James P and Bierbaum, Matthew K and Dahmen, Karin A and Goodrich, Carl P and Greer, Julia R and Hayden, Lorien X and Kent-Dobias, Jaron P and Lee, Edward D and Liarte, Danilo B and Ni, Xiaoyue and others},
  journal={Annu. Rev. Mater. Res.},
  volume={47},
  number={1},
  pages={217--246},
  year={2017},
  publisher={Annual Reviews}
}

@article{jung2025roadmap,
  title={Roadmap on machine learning glassy dynamics},
  author={Jung, Gerhard and Alkemade, Rinske M and Bapst, Victor and Coslovich, Daniele and Filion, Laura and Landes, Fran{\c{c}}ois P and Liu, Andrea J and Pezzicoli, Francesco Saverio and Shiba, Hayato and Volpe, Giovanni and others},
  journal={Nat. Rev. Phys.},
  volume={7},
  number={2},
  pages={91--104},
  year={2025},
  publisher={Nature Publishing Group UK London}
}

@article{lin2014scaling,
  title={Scaling description of the yielding transition in soft amorphous solids at zero temperature},
  author={Lin, Jie and Lerner, Edan and Rosso, Alberto and Wyart, Matthieu},
  journal={Proc. Natl. Acad. Sci.},
  volume={111},
  number={40},
  pages={14382--14387},
  year={2014},
  publisher={National Academy of Sciences}
}

@article{fisher1998collective,
  title={Collective transport in random media: from superconductors to earthquakes},
  author={Fisher, Daniel S},
  journal={Phys. Rep.},
  volume={301},
  number={1-3},
  pages={113--150},
  year={1998},
  publisher={Elsevier}
}

@article{reichhardt2016depinning,
  title={Depinning and nonequilibrium dynamic phases of particle assemblies driven over random and ordered substrates: a review},
  author={Reichhardt, Charles and Reichhardt, CJ Olson},
  journal={Rep. Prog. Phys.},
  volume={80},
  number={2},
  pages={026501},
  year={2016},
  publisher={IOP Publishing}
}

@book{berthier2011dynamical,
  title={Dynamical heterogeneities in glasses, colloids, and granular media},
  author={Berthier, Ludovic and Biroli, Giulio and Bouchaud, Jean-Philippe and Cipelletti, Luca and van Saarloos, Wim},
  volume={150},
  year={2011},
  publisher={OUP Oxford}
}

@article{chen2012topological,
  title={Topological rearrangements and stress fluctuations in quasi-two-dimensional hopper flow of emulsions},
  author={Chen, Dandan and Desmond, Kenneth W and Weeks, Eric R},
  journal={Soft Matter},
  volume={8},
  number={40},
  pages={10486--10492},
  year={2012},
  publisher={Royal Society of Chemistry}
}

@article{campbell2010jamming,
  title={Jamming and unjamming of concentrated colloidal dispersions in channel flows},
  author={Campbell, Andrew I and Haw, Mark D},
  journal={Soft Matter},
  volume={6},
  number={19},
  pages={4688--4693},
  year={2010},
  publisher={Royal Society of Chemistry}
}

@article{isa2009velocity,
  title={Velocity oscillations in microfluidic flows of concentrated colloidal suspensions},
  author={Isa, Lucio and Besseling, Rut and Morozov, Alexander N and Poon, Wilson C},
  journal={Physical review letters},
  volume={102},
  number={5},
  pages={058302},
  year={2009}
}

@article{chehata2003dense,
  title={Dense granular flow around an immersed cylinder},
  author={Chehata, D and Zenit, R and Wassgren, CR},
  journal={Physics of fluids},
  volume={15},
  number={6},
  pages={1622--1631},
  year={2003},
  publisher={American Institute of Physics}
}

@article{tokpavi2009experimental,
  title={Experimental study of the very slow flow of a yield stress fluid around a circular cylinder},
  author={Tokpavi, Dodji L{\'e}agnon and Jay, Pascal and Magnin, Albert and Jossic, Laurent},
  journal={Journal of Non-Newtonian Fluid Mechanics},
  volume={164},
  number={1-3},
  pages={35--44},
  year={2009},
  publisher={Elsevier}
}

@article{zisis2002viscoplastic,
  title={Viscoplastic flow around a cylinder kept between parallel plates},
  author={Zisis, Th and Mitsoulis, E},
  journal={Journal of non-newtonian fluid mechanics},
  volume={105},
  number={1},
  pages={1--20},
  year={2002},
  publisher={Elsevier}
}

@article{mousavi2025elasto,
  title={Elasto-visco-plastic flows in benchmark geometries: II. Flow around a confined cylinder},
  author={Mousavi, Milad and Dimakopoulos, Yannis and Tsamopoulos, John},
  journal={Journal of Non-Newtonian Fluid Mechanics},
  volume={336},
  pages={105384},
  year={2025},
  publisher={Elsevier}
}

@article{Durian1997,
  title={Bubble-scale model of foam mechanics: Melting, nonlinear behavior, and avalanches},
  author={Durian, D. J.},
  journal={Phys. Rev. E},
  volume={55},
  pages={1739--1751},
  year={1997}
}

@article{Okuzono1995,
  title={Intermittent flow behavior of random foams: A computer experiment on foam rheology},
  author={T. Okuzono and K. Kawasaki},
  journal={Phys. Rev. E},
  volume={51},
  pages={1246--1253},
  year={1995}
}

@article{cantat2011gibbs,
  title={Gibbs elasticity effect in foam shear flows: a non quasi-static 2D numerical simulation},
  author={Cantat, Isabelle},
  journal={Soft Matter},
  volume={7},
  number={2},
  pages={448--455},
  year={2011},
  publisher={Royal Society of Chemistry}
}

@article{Kern2004,
  title={Two-dimensional viscous froth model for foam dynamics},
  author={N. Kern and D. Weaire and A. Martin and S. Hutzler and S. J. Cox},
  journal={Phys. Rev. E},
  volume={70},
  pages={041411},
  year={2004}
}

@article{Morse1993,
  title={Droplet elasticity in weakly compressed emulsions},
  author={D. C. Morse and T. A. Witten},
  journal={Europhys. Lett.},
  volume={22},
  pages={549--555},
  year={1993}
}

@article{Hohler2019,
  title={Can liquid foams and emulsions be modeled as packings of soft elastic particles?},
  author={R. Höhler and D. Weaire},
  journal={Adv. Colloid Interface Sci.},
  volume={263},
  pages={19--37},
  year={2019}
}

@article{Denkov2008,
  title={Viscous friction in foams and concentrated emulsions under steady shear},
  author={N. D. Denkov and S. Tcholakova and K. Golemanov and K. P. Ananthapadmanabhan and A. Lips},
  journal={Phys. Rev. Lett.},
  volume={100},
  pages={138301},
  year={2008}
}

@article{Raufaste2009,
  title={Dissipation in quasi-two-dimensional flowing foams},
  author={C. Raufaste and A. Foulon and B. Dollet},
  journal={Phys. Fluids},
  volume={21},
  pages={053102},
  year={2009}
}

@article{BenSalem2013,
  title={Response of a two-dimensional liquid foam to air injection: swelling rate, fingering and fracture},
  author={I. {Ben Salem} and I. Cantat and B. Dollet},
  journal={J. Fluid Mech.},
  volume={714},
  pages={258--282},
  year={2013}
}

\end{document}